\documentclass[preprint,pra,aps,nofootinbib,longbibliography]{revtex4-2}

\usepackage{amsmath,amssymb,amstext}
\usepackage[pdftex]{graphicx} 
\usepackage{tabularx}
\usepackage{newtxtext,newtxmath}
\usepackage{bm}
\usepackage{mathrsfs}
\usepackage{fancyhdr}
\usepackage{booktabs}
\usepackage{longtable}
\usepackage{multirow}

\usepackage[pdftex,pagebackref=false]{hyperref} 
\hypersetup{
    unicode=false,          
    pdffitwindow=false,     
    pdfstartview={FitH},    
    pdfnewwindow=true,      
    colorlinks=true,        
    linkcolor=blue,         
    citecolor=green,        
    filecolor=magenta,      
    urlcolor=cyan           
}

\let\origdoublepage\cleardoublepage
\newcommand{\clearemptydoublepage}{%
  \clearpage{\pagestyle{empty}\origdoublepage}}
\let\cleardoublepage\clearemptydoublepage

\allowdisplaybreaks

\newcommand{\Lim}[1]{\raisebox{0.5ex}{\scalebox{1.0}{$\displaystyle \lim_{#1}\;$}}}

\begin{document}

\title{On the Origins of Spontaneous Spherical Symmetry-Breaking in Open-Shell Atoms Through Polymer Self-Consistent Field Theory}
\author{Phil A. LeMaitre}
\email{plemaitr@uwaterloo.ca}
\affiliation{Department of Physics \& Astronomy, University of Waterloo, 200 University Avenue West, Waterloo, Ontario, Canada N2L 3G1}
\author{Russell B. Thompson}
\affiliation{Department of Physics \& Astronomy and Waterloo Institute for Nanotechnology, University of Waterloo, 200 University Avenue West, Waterloo, Ontario, Canada N2L 3G1}
\date{\today}

\begin{abstract}
An alternative approach to density functional theory based on self-consistent field theory for ring polymers is applied to neutral atoms hydrogen to neon in their ground-states. The spontaneous emergence of atomic shell structure and spherical symmetry-breaking of the total electron density is predicted by the model using ideas of polymer excluded-volume between pairs of electrons to enforce the Pauli-exclusion principle, and an exact electron self-interaction correction. The Pauli potential is approximated by neglecting inter-atomic correlations along with other types of correlations and comparisons to Hartree-Fock theory are made, which also ignores correlations. The model shows excellent agreement with Hartree-Fock theory for the atomic binding energies and density profiles of the first six elements, providing exact matches for the elements hydrogen and helium. The predicted shell structure starts to deviate significantly past the element neon and spherical symmetry-breaking is first predicted to occur at carbon instead of boron. The self-consistent field theory energy functional which describes the model is decomposed into thermodynamic components to trace the origin of spherical symmetry-breaking. It is found to arise from the electron density approaching closer to the nucleus in non-spherical distributions, which lowers the energy despite resulting in frustration between the quantum kinetic energy, electron-electron interaction, and the Pauli exclusion interaction. The symmetry-breaking effect is found to have minimal impact on the binding energies, which suggests that the spherical-averaging approximation used in previous work is physically reasonable when investigating atomic systems. The pair density contour plots display behaviour similar to polymer macro-phase separation, where individual electron pairs occupy single lobe structures that together form a dumbbell shape analogous to the 2p orbital shape. It is further shown that the predicted densities satisfy known constraints and produce the same total electronic density profile that is predicted by other formulations of quantum mechanics. 
\end{abstract}   

\cleardoublepage

\maketitle

\section{Introduction}
\label{1}

Two features of atomic systems that emerge from quantum theory are the presence of highly inhomogeneous shell structure and the spontaneous breaking of spherical symmetry. These two attributes play an important role in determining atomic size and facilitating chemical bonding. An intuitive understanding of the mechanism behind them would be helpful in considerations of molecules and bulk materials. On one hand, quantum superposition arguments can be made in support of the point of view that all isolated atoms, even open shell atoms, are spherically symmetric \cite{Cohen1965, gil1972}. On the other hand, the concept of an ``isolated'' atom is purely theoretical, and operationally, it is nowadays accepted that  open-shell atoms used as building blocks for molecular calculations are not generally treated as have spherically symmetric electron densities \cite{Fertig2000}. In particular, the recent density functional theory (DFT) study of open-shell atoms by Chowdhury and Perdew \cite{Chowdhury_Perdew2021} has examined the implications of spherical symmetry-breaking for molecular bonding and atomization energies.

An alternative to the Kohn-Sham (KS) DFT used by Chowdhury and Perdew \cite{Chowdhury_Perdew2021} is polymer self-consistent field theory (SCFT). Based on a quantum-classical isomorphism introduced by Feynman \cite{Feynman1953AtomicTO, Chandler_Wolynes1981}, classical statistical mechanics is used to represent quantum particles as ring ``polymers'', that is, extended non-point like objects, embedded in an extra thermal dimension \cite{thompson2019.article}. SCFT has been shown to be equivalent to KS-DFT, assuming the Pauli-exclusion principle holds \cite{thompson2019.article}, and incorporates all quantum effects \cite{Thompson2022} in terms of classical quantities which enables explanations of quantum phenomena more in tune with classical intuition. The use of a Pauli potential and lack of explicit orbitals means that SCFT is related to orbital-free (OF) DFT, and as is typical in OF-DFT, the exclusion principle can be incorporated using approximations \cite{thompson2020.article}. SCFT predicts molecular bonding \cite{sillaste_thompson.article}, atomic shell structure \cite{LeMaitre_Thompson_2022}, includes temperature dependence \cite{thompson2019.article}, and can be related to dynamical quantum mechanics \cite{Thompson2022}. SCFT also has foundational implications due to its classical ensemble formulation \cite{Thompson2022}. Like DFT, SCFT is based on an energy functional, and since the quantities are real and classical, it can be decomposed into thermodynamic components in order to explain the origins of predicted structures \cite{Matsen_Bates1997}. 

The purpose of this paper is to show that SCFT spontaneously predicts shell structure and spherical symmetry-breaking in isolated atoms, and to give the thermodynamic explanations for non-spherical ground-state structures. We partition the SCFT free energy functional for the electron density into components: the translational entropy contribution, polymer configurational entropy contribution (equivalent to the quantum kinetic energy), and the internal energy which includes electron-electron, electron-nucleus, and Pauli-exclusion interactions. We find that for the atoms we studied from hydrogen to neon, all thermodynamic contributions cause the free energy to increase for spherical symmetry-breaking except the electron-nucleus potential. In other words, the overall energy is reduced when the electrons can move closer to the nucleus by breaking spherical symmetry and this more than compensates for all other factors which would tend to increase the energy.

Neutral atoms in their ground-state were studied in a previous work \cite{LeMaitre_Thompson_2022} using a spherical-averaging approximation, which amounts to representing only the radial part of the electron density, as well as the adoption of partial atomic shell information. These approximations were useful to see how well the model could replicate atomic trends for a large range of elements, most importantly if it could produce the proper shell structure. We will generalize that work by restoring the full angular distribution of the electron density. We organize the paper as follows. In section \ref{2}, the conceptual basis of the theory is outlined and a new derivation of the model equations is given. The fields in the model are introduced, the free energy components are given, and the section ends with a discussion about the numerical methods used. The atomic binding energies for the elements hydrogen to neon and the angular electron density contour plots for the elements boron to neon are presented in section \ref{3}, along with tables of data for the free energy components and various density constraints that the model satisfies with a proof of the equivalence with KS-DFT. The Discussion in section \ref{4} demonstrates that the external potential is responsible for the spherical symmetry breaking. The paper concludes in section \ref{5} and some future research directions are discussed.

\section{Theory}
\label{2}

The pioneering work of Feynman in developing the path integral formulation of quantum theory led to the deduction of the isomorphism between quantum mechanics and the statistical mechanics of ring-like molecules, using a Wick rotation of time $t=-i\hbar\beta$ to a parameter $\beta$ \cite{Feynman1953AtomicTO}. The Wick rotation essentially allows one to transform a dynamics problem into a statics problem in one higher dimension, where temporal variables are typically exchanged for spatial variables \cite{lancaster2014}. Part of the intuition behind why the Wick rotation takes quantum mechanics to the statistical mechanics of ring-like molecules stems from the nature of the trace operator in the partition function, which is a sum of the Boltzmann factors over configurations starting and ending at the same position. This cyclic aspect is crucial to the interpretation, especially after having transferred to the path integral picture, because the path integral allows one to see that each of the trajectories followed through the imaginary time configuration parameter space correspond to the different possible arrangements of quantum particles comprising a system of quantum particles \cite{Feynman1953AtomicTO,Chandler_Wolynes1981}. The correspondence with the mathematics of polyatomic fluids noted by Chandler \cite{Chandler_Wolynes1981} lends itself to label each of the quantum particles in the system as atoms comprising a molecule. The probabilistic uncertainty in the position and momenta of the quantum system are then manifested as the collection of system arrangements associated to a given configuration. The connection to polymer SCFT is then made clear through the insights of Matsen \cite{matsen.incollection}. The periodicity of the imaginary time parameter $\beta$ also happens to be a part of one of the conditions to be a Kubo-Martin-Schwinger (KMS) state, which is a general notion of being in a thermal state \cite{haag1967}. Although the mathematics is essentially identical between quantum mechanical and polymeric systems, the interpretation of the fundamental constituents and the mechanisms that govern their behaviour has changed dramatically. In the quantum mechanical case, the time evolution of the system could trace out many different paths, forming a distribution of them that expresses the uncertainty in which path will be followed. In the statistical mechanical case, the system could explore many different energy configurations, forming a distribution of them where thermal fluctuations represent the uncertainty in which configuration will be chosen. A set of postulates has been proposed to bridge between finite-temperature quantum DFT and ring polymer SCFT \cite{Thompson2022}.

\begin{figure}[h]
\centering
\includegraphics[height=10cm, width=7cm]{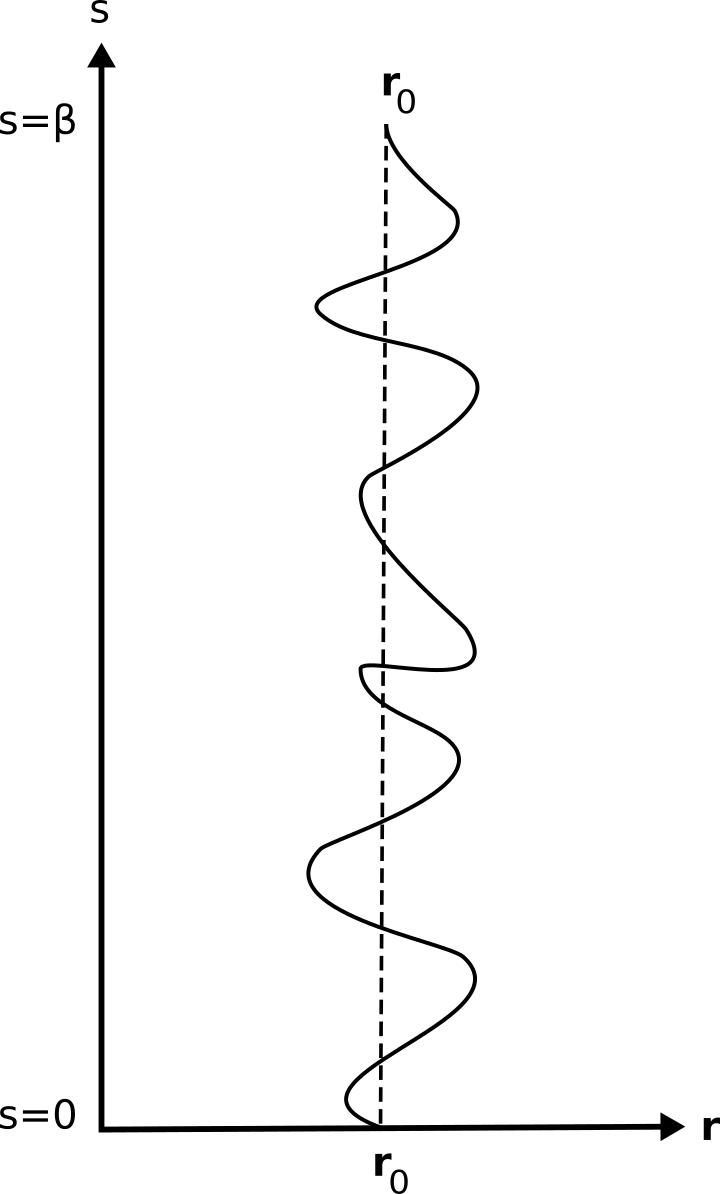}
\caption{A visual schematic of a quantum particle trajectory in $s$ parameter space, where the horizontal axes are the spatial coordinates and the vertical axes are the $s$ coordinates. The quantum particle starts out in a position $\bm{r}_0$ at $s=0$ and follows the $s$-space trajectory to the same starting position but at $s=\beta$, representing a ``ring''.} \label{poly_contour}
\end{figure}

In the static case considered here, there are two postulates needed, in addition to those from statistical mechanics, to describe quantum many-body systems of Fermions: 1) ``pairs'' (i.e. 1-2) of quantum particles are classically modelled as stochastic chains in a 4-dimensional thermal space and 2) the chains have excluded-volume with respect to 4-dimensional thermal space. The first postulate can be seen to be equivalent to the Heisenberg uncertainty principle \cite{thompson2019.article,thompson2020.article}, and the introduction of the term ``pairs'' is to account for Fermion spin, while the second postulate is a (conjectured) statement of the Pauli-exclusion principle. Similar to ring polymer SCFT, the base entities in this theory are fundamentally distinguishable from each other, so the notion of particle ``pairs'' with higher-dimensional excluded-volume is the mechanism that emulates particle statistics from quantum statistical mechanics. To recover the standard electron density, field, and electron number that appear in DFT, we simply sum up all of the pairs, which we denote with a Greek index. The two postulates in combination with the usual approximations used in the study of many-body systems (i.e. Born-Oppenheimer approximation, point-particle nuclei) are then enough to derive the expression for the partition function of the system. Working in the canonical ensemble for simplicity, with $N$-body potential $U$, the $N$-body partition function $Q_N$ can be expressed as a path integral in configuration space as
\begin{align}
Q_N(\beta) =  \prod^N_i\int\mathrm{d}\bm{r}_i\int\mathcal{D}[\bm{r}_i] e^{-\int^{\beta}_0 \mathrm{d}\tau \left[\sum_j \frac{m}{2\hbar^2}\left|\frac{\mathrm{d}\bm{r}_j(\tau)}{\mathrm{d}\tau}\right|^2+U[\hat{n}](\bm{r}_1(\tau), \ldots, \bm{r}_N(\tau))\right]} \label{expec_bolt}
\end{align}
where the parameter $\beta$ is the Lagrange multiplier that ensures the expectation value of the free energy remains constant, which will end up being the reciprocal thermal energy $\beta=1/k_BT$ (where $k_B$ is Boltzmann's constant and $T$ is the temperature) due to the KMS condition \cite{haag1967}; $\bm{r}_i(s)$ is the parametrized curve representing the $i$th quantum particle as depicted in figure \ref{poly_contour}, with the parameter $s$ running from $0$ (a high classical temperature) to $\beta$ (a lower temperature); and the $N$-body potential $U$ is expressed as a functional of an electron density operator $\hat{n}(\bm{r})$, which is defined to be
\begin{align}
\hat{n}(\bm{r}) = \sum_\mu \hat{n}_\mu(\bm{r}) = \sum_\mu\sum^{N_\mu}_i \delta(\bm{r}-\bm{r}_i)\,. \label{elec_denop}
\end{align}    
$U[\hat{n}]$ can be re-expressed in terms of the fields $\mathcal{N}_\mu(\bm{r})$ using the functional Dirac delta $\delta[\mathcal{N}_\mu-\hat{n}_\mu]$, which can in-turn be expressed in terms of its functional Fourier transform representation with respect to conjugate fields $W_\mu(\bm{r})$, allowing eqn. \ref{expec_bolt} to be expressed as 
\begin{align}
Q_N(\beta) &= \prod_\mu\int \int \mathcal{D}[\mathcal{N}_\mu] \mathcal{D}[W_\mu]e^{-\beta U[\mathcal{N}] + \beta\int \mathrm{d}\bm{r}' W_\mu(\bm{r}')\mathcal{N}_\mu(\bm{r}')}\prod^{N_\mu}_i\int\mathrm{d}\bm{r}_i \nonumber \\ 
&\qquad \qquad \times\int\mathcal{D}[\bm{r}_i] e^{-\int^\beta_0 \mathrm{d}\tau \left[ \frac{m}{2\hbar^2}\left|\frac{\mathrm{d}\bm{r}_i(\tau)}{\mathrm{d}\tau}\right|^2 + \int \mathrm{d}\bm{r}' W_\mu(\bm{r}')\hat{n}_\mu(\bm{r}')\right]}\,. \label{part1}
\end{align}
The $\hat{n}$ operator implicitly carries the $N$ coordinate dependencies with it, so inserting eqn. \ref{elec_denop} into eqn. \ref{part1} allows one to see, after some manipulations, that the argument of the exponential in the configuration path integral is now completely separable into $N$ one-body terms, producing a product of $N$ separable path integrals. The configuration integrals can then be evaluated one at a time with the result being $N$ identical terms. The $N$-body partition function $Q_N(\beta)$ can then finally be expressed as
\begin{align}
Q_N(\beta) = \prod_\mu\int \int \mathcal{D}[\mathcal{N}_\mu] \mathcal{D}[W_\mu]\,e^{-\beta F[\mathcal{N}_\mu, W_\mu]} \label{part_fin}
\end{align}
where
\begin{align}
F[\mathcal{N}_\mu, W_\mu] = -\frac{1}{\beta} N_\mu\ln (Q_\mu[W](\beta))+U[\mathcal{N}_\mu]- \int \mathrm{d}\bm{r}' W_\mu(\bm{r}')\mathcal{N}_\mu(\bm{r}') \label{free_func}
\end{align}
and $Q_\mu[W](\beta)$ is a single-pair partition function that we have defined according to the expression 
\begin{align}
Q_\mu[W](\beta) =  \int \mathrm{d}\bm{r}\, q_\mu(\bm{r}, \bm{r}, \beta)  \label{1part}
\end{align}
where $q_\mu(\bm{r}, \bm{r}, \beta)$ represents the propagation of a single pair from initial position $\bm{r}$ at $s=0$ to final position $\bm{r}$ at $s=\beta$. The single-pair propagator $q_\mu(\bm{r}, \bm{r}', s)$ can be expressed as a path integral: 
\begin{align}
q_\mu(\bm{r}, \bm{r}', s) = \mathcal{A}\int \mathcal{D}[\bm{r}] e^{-\int^s_0 \mathrm{d}\tau  \left[\frac{m}{2\hbar^2}\left|\frac{\mathrm{d}\bm{r}(\tau)}{\mathrm{d}\tau}\right|^2 + W_\mu(\bm{r}(\tau))\right]} \label{prop1}
\end{align}
where $\mathcal{A}$ is a formally infinite normalization constant coming from the kinetic degrees of freedom, whose value we shall not be concerned with since it will not appear in any of the quantities of interest. As a propagator, it can be shown that $q_\mu(\bm{r}, \bm{r}', s)$ equivalently satisfies the modified diffusion equation
\begin{align}
\frac{\partial q_\mu(\bm{r}, \bm{r}', s)}{\partial s} = -H^{\text{eff}}_\mu q_\mu(\bm{r}, \bm{r}', s) = \frac{\hbar^2}{2m}\nabla^2q_\mu(\bm{r}, \bm{r}', s) - W_\mu(\bm{r})q_\mu(\bm{r}, \bm{r}', s) \label{diffuse}
\end{align}
with initial condition $q_\mu(\bm{r}, \bm{r}', 0) = \delta(\bm{r}-\bm{r}')$, which makes it possible to evaluate $Q_\mu[W](\beta)$ and thus $F[\mathcal{N}, W]$ exactly. It is worth pointing out that the Hamiltonian $H_{\text{eff}}$ above is the same Hamiltonian appearing in the Kohn-Sham equation from KS-DFT, a fact we use to prove the equivalence of the two theories in appendix \ref{AppendixC}. Although it is possible to analytically evaluate the functional eqn. \ref{free_func}, we are not so fortunate with eqn. \ref{part_fin}, whose path integrals are too unwieldy to perform exact calculations with to get the free energy \cite{Matsen_Schick1994.article}. However, eqn. \ref{free_func} plays the same role that the action does in the real-time quantum mechanical path integral, so a solution can be sought which extremizes $F[\mathcal{N}, W]$ by setting its first variation to zero and then approximating the integrand with the extremum of $F$. The free energy can then be calculated from $F[n, w]$, where $n$ and $w$ are the mean-fields for which the functional $F$ has a saddle point \cite{matsen.incollection,Matsen_Schick1994.article,Kim_Yang_Lee2012}. We can further justify the preservation of exactness in the model from varying eqn. \ref{free_func}, since any neglected higher-order contributions can be packaged into the unknown functional $U$, whose approximations occupy a large portion of current DFT research \cite{Jones2015,witt_del_rio_dieterich_carter2018,Karasiev_Trickey2012}. The total kinetic energy functional $K[n, w]$ can also be calculated from the expectation value of the kinetic term in the many-body Hamiltonian, after some manipulation, to be
\begin{align}
K[n, w] &= -\sum^N_j\bigg\langle \frac{\hbar^2}{2m}\nabla^2_j\bigg\rangle = -\frac{\hbar^2}{2m}\sum_\mu\sum^N_j\frac{1}{Q_\mu[w](\beta)} \int \mathrm{d}\bm{r}_j  \nabla^2_{\bm{r}'_j}q_\mu(\bm{r}_j, \bm{r}'_j, \beta)\bigg|_{\bm{r}'_j=\bm{r}_j} \,. \label{kin_func}
\end{align}
Proceeding from the variational principle outlined above, using the path integral form of the single-pair propagator, the mean-field density $n_\mu(\bm{r}, \beta)$ and field $w_\mu(\bm{r}, \beta)$ corresponding to each pair are found to be 
\begin{align}
w_\mu(\bm{r}, \beta) = \frac{\delta U[n]}{\delta n_\mu(\bm{r}, \beta)} \ \ \text{and} \ \ n_\mu(\bm{r}, \beta) = \frac{N_\mu}{Q_\mu[w](\beta)}q_\mu(\bm{r}, \bm{r}, \beta)\,.\label{field_dens}
\end{align}  
The potential $U[n]$ is the only remaining quantity yet to be specified, which will give us the expressions for the fields $w_\mu(\bm{r}, \beta)$ experienced by each pair in the system, and finally the expression for the free energy $F[n, w]$. In our model, the electron pairs experience four fields in the vicinity of the atomic nucleus: the Coulomb field between the nucleus and the electron pairs $w_\mu^{e-n}(\bm{r}, \beta)$, the Coulomb field between the electrons $w_\mu^{e-e}(\bm{r}, \beta)$, and the exchange field between electron pairs $w_\mu^{x}(\bm{r}, \beta)$ representing two separate fields. The first of these two fields is the electron self-interaction field $w_\mu^{sic}(\bm{r}, \beta)$, which corrects for the interaction of the electron with its own field that is not accounted for in the electron-electron Coulomb field $w_\mu^{e-e}(\bm{r}, \beta)$; the self-interaction correction introduced in prior work \cite{LeMaitre_Thompson_2022} is employed in this work as well. The second of these two fields, as is commonplace in all OF-DFT approaches, is the Pauli-exclusion field $w_\mu^P(\bm{r}, \beta)$, which accounts for the repulsion felt by electron pairs with the same configuration attempting to occupy the same location at the same (imaginary) time, as stipulated by the Pauli-exclusion principle. As will be discussed later, the Pauli-exclusion field used in this work accounts for some exchange effects but does not account for correlations.

The electron-nucleus potential $U_{e-n}[n]$ takes the form
\begin{align}
U_{e-n}[n] = -\int \int \mathrm{d}\bm{r}\, \mathrm{d}\bm{r}'\, n(\bm{r}, \beta)\frac{\rho_{\text{nuc}}(\bm{r}')}{\left|\bm{r}-\bm{r}' \right|}\,,
\end{align}
where $\rho_{\text{nuc}}(\bm{r})$ is the nuclear density, which we take to be $\rho_{\text{nuc}}(\bm{r}) = N\delta(\bm{r})$. The electron-nucleus field for each pair $w^{e-n}_\mu(\bm{r}, \beta)$ is then found to be 
\begin{align}
w^{e-n}_\mu(\bm{r}, \beta) = - \int \mathrm{d}\bm{r}'\, \frac{\rho_{\text{nuc}}(\bm{r}')}{\left|\bm{r}-\bm{r}' \right|} = -\frac{N}{|\bm{r}|}\,. \label{wen}
\end{align} 
The potential due to electron-electron Coulomb-type interactions $U_{e-e}[n]$ is similarly given by 
\begin{align}
U_{e-e}[n] = \frac{1}{2}\int \int \mathrm{d}\bm{r}\, \mathrm{d}\bm{r}'\, n(\bm{r}, \beta)\frac{n(\bm{r}', \beta)}{\left|\bm{r}-\bm{r}' \right|}\,,
\end{align}
and the electron-electron field for each pair $w^{e-e}_\mu(\bm{r}, \beta)$ is then found to be 
\begin{align}
w^{e-e}_\mu(\bm{r}, \beta) = \int \mathrm{d}\bm{r}' \frac{n(\bm{r}', \beta)}{|\bm{r}'-\bm{r}|}\,. \label{wee}
\end{align}
Both eqn. \ref{wen} and eqn. \ref{wee} indicate that each pair experiences the exact same electron-electron and electron-nucleus field. The total field experienced by each pair will however not be the same, thanks to the exchange effects introduced by the other two fields.

Following previous work \cite{thompson2020.article}, the closest classical analogue of the Pauli-exclusion principle is the notion of excluded-volume, which in polymer SCFT, is often implemented as a Dirac delta energy penalty for overlapping polymer segments. If we are to be truly faithful to the exclusion principle however, the energy penalty should be for overlapping polymer segments from \textit{different} polymer contours representing pairs of quantum particles to account for spin. Since the position along the polymer contour is parametrized by a parameter $s$, the energy penalty due to overlapping polymer contours occurs only for contours at the same value of $s$. Recall from the quantum-classical correspondence that the parameter $s$ can be interpreted as an imaginary time, so the Pauli-exclusion repulsion is akin to a particle pair feeling the excluded-volume repulsion when at the same place and (imaginary) time as another pair \cite{thompson2020.article}. This idea is difficult to implement in practice however, so we approximate it by projecting out the degrees of freedom from the $s$ parameter space, which effectively amounts to imposing the excluded-volume energy penalties for every value of $s$. The downside to this approximation is that it ignores the inter-contour correlations and will clearly overestimate the excluded-volume felt between the pairs \cite{thompson2020.article}. The approximate Pauli-exclusion potential is then given by
\begin{align}
U_P[n] = \frac{1}{2g_0}\sum_{\substack{\mu,\nu \\ \mu \neq \nu}} \int \int \mathrm{d}\bm{r}\mathrm{d}\bm{r}' n_\mu(\bm{r}, \beta)\delta(\bm{r}-\bm{r}')n_\nu(\bm{r}', \beta) = \frac{1}{2g_0}\sum_{\substack{\mu,\nu \\ \mu \neq \nu}} \int \mathrm{d}\bm{r} n_\mu(\bm{r}, \beta)n_\nu(\bm{r}, \beta) \label{Paulipot_approx}
\end{align}
where $g_{0}$ is a constant with the same units as a density of states \cite{thompson2020.article}. In principle, since the excluded-volume interaction is independent of the system under study, $g_0$ should be a universal constant whose value can be determined by comparing the Pauli potential for a very simplistic system (e.g. a uniform gas with only excluded-volume interactions) to experimental results. However, because the Pauli potential is being approximated in this work, $g_0$ is taken to be arbitrary and we choose its value once for all calculations; the value chosen and how it was chosen will be discussed in the Results section \ref{3}. The approximate form of the Pauli-exclusion field for each pair $w^{P}_\mu(\bm{r}, \beta)$ is then calculated as 
\begin{align}
w^{P}_\mu(\bm{r}, \beta) = \frac{1}{g_0} \sum_{\substack{\gamma \\ \gamma \neq \mu}}n_\gamma(\bm{r}, \beta)\,. \label{wpauli}
\end{align}
In previous work \cite{LeMaitre_Thompson_2022}, some constraints on the exact form of the Pauli-exclusion field were given which allowed for the accuracy of the approximate expression eqn. \ref{wpauli} to be assessed. The relevant constraints were as follows:
\begin{align}
&w^P(\bm{r}, \beta) \geq 0\ \ , \ \ \Lim{|\bm{r}|\rightarrow \infty} w^P(\bm{r}, \beta) = 0\ \ , \ \ \int \mathrm{d}\bm{r}\, w^P(\bm{r}, \beta)n(\bm{r}, \beta) < \infty\ \ , \ \ w^P(\bm{r}, \beta) = 0 \ \ \text{for} \ \ N=2\ \ , \nonumber \\[1.5ex]  &w^P[\lambda^3n](\lambda\bm{r}, \beta) = \lambda^2w^P[n](\lambda\bm{r}, \beta) \label{pauli_cons}
\end{align}
where in the last criterion $\lambda$ is a scale factor and the functional dependence of the field on the density has been explicitly reinstated \cite{Levy_Ou-Yang1988.article}. We found that all but the last of the constraints in eqn. \ref{pauli_cons} were satisfied by eqn. \ref{wpauli}, with $w^P(\bm{r}, \beta)$ overestimating the excluded-volume interactions by precisely the amount required to fulfill the last constraint in eqn. \ref{pauli_cons}; a point also discussed in a previous work \cite{thompson2020.article}.

The electron self-interaction field $w_\mu^{\text{sic}}(\bm{r}, \beta)$ used in this work was first introduced in reference \cite{LeMaitre_Thompson_2022}, and is essentially a Fermi-Amaldi self-interaction correction applied to each particle pair; see reference \cite{LeMaitre_Thompson_2022} for further discussion. It has the form
\begin{align}
w_\mu^{\text{sic}}(\bm{r}, \beta) = -\frac{1}{N_\mu}\int \mathrm{d}\bm{r}'\frac{n_\mu(\bm{r}', \beta)}{|\bm{r}-\bm{r}'|}\label{self_int}
\end{align}
where the corresponding potential $U_{\text{sic}}[n]$ is simply
\begin{align}
U_{\text{sic}}[n] = -\sum_\mu\frac{1}{2N_\mu}\int \int \mathrm{d}\bm{r}\mathrm{d}\bm{r}'\frac{n_\mu(\bm{r}', \beta)n_\mu(\bm{r}, \beta)}{|\bm{r}-\bm{r}'|}\,. \label{self_int_pot}
\end{align}
In this form, eqn. \ref{self_int} directly preserves the desirable qualities of the original Fermi-Amaldi electron self-interaction correction for hydrogen and helium \cite{ayers_morrison_parr.article}. Furthermore, because eqn. \ref{self_int} acts on electron pairs and $N_\mu = \{1, 2\}$, then eqn. \ref{self_int} effectively accounts for the self-interaction of every electron in the atomic system. Those familiar with the Hartree-Fock model will immediately recognize that our model with eqn. \ref{self_int} is \textit{identical} to the situation in the Hartree-Fock model, which models exchange effects exactly. However, because we are approximating the Pauli-exclusion field by projecting out the degrees of freedom from the $s$ parameter space, our model in its current implementation will not reproduce the precise binding energies predicted by Hartree-Fock theory. This is because our Pauli-exclusion field overestimates the excluded-volume felt by the polymer contours in the $s$-parameter space, hence electron pairs feel too much repulsion between each other and the electron shells will be too distant from their neighbours, raising the free energy substantially in some cases. The Hartree-Fock model on the other hand, is a wavefunction-based model, so the Pauli-exclusion effect is automatically encoded into the wavefunction due to the spin-statistics theorem \cite{szabo2012modern}. Implementing the exact expression for the Pauli potential would then allow our model to coincide exactly with Hartree-Fock theory.

As was mentioned previously, electron pairs are taken to be ring polymers embedded in a 3+1-dimensional thermal space under the influence of a potential $U$. The ring polymers are confined to explore the thermal space according to this potential, which they can do using two different mechanisms: translation and configuration. The translational degrees of freedom refer to the three dimensional motion of the polymer as a whole, while the configurational degrees of freedom refer to how the polymer confirmations change while holding one point of the polymer fixed in space. Each of these mechanisms has an entropy associated to them, which we will denote as $S_t$ and $S_c$, respectively. The behaviour of the polymer can then be explained by looking at the competition between the degrees of freedom encoded in the entropies and those restricted by the potential $U$. Therefore, by rephrasing the free energy per pair $F[n_\mu, w_\mu]$ in terms of these entropies, we can exactly describe the process in the polymer picture by which the electrons surrounding the atomic nucleus would break spherical-symmetry; and with the notion of pairs, we can pinpoint exactly which pairs affect this change. Following previous work \onlinecite{thompson2019.article} and reference \onlinecite{Matsen_Bates1997}, the free energy per pair $F[n_\mu, w_\mu]$ can be re-expressed as
\begin{align}
F[n_\mu, w_\mu] &= -\frac{1}{\beta} \int \mathrm{d}\bm{r}' n_\mu(\bm{r}', \beta)\ln{\left(\frac{n_\mu(\bm{r}', \beta)}{N_\mu}\right)}+U[n_\mu] \nonumber \\[1.5ex]
&\quad+\frac{1}{\beta}\int \mathrm{d}\bm{r}' n_\mu(\bm{r}', \beta)\left[\ln{\left(q_\mu(\bm{r}', \bm{r}', \beta)\right)} + \beta w_\mu(\bm{r}', \beta)\right] \label{free_func_ent}
\end{align}
where we can identify the last term with the free energy contribution coming from the configurational entropy $S_c[n_\mu, w_\mu]$ \footnote{This configurational entropy expression corrects a typo in reference \onlinecite{thompson2019.article} which is missing a factor of $-\beta$.} and the first term with the translational entropy $S_t[n_\mu]$, which we write as
\begin{align}
S_c[n_\mu, w_\mu] &= -\int \mathrm{d}\bm{r}' n_\mu(\bm{r}', \beta)\left[\ln{\left(q_\mu(\bm{r}', \bm{r}', \beta)\right)} + \beta w_\mu(\bm{r}', \beta)\right]\\[1.5ex]
S_t[n_\mu] &= \int \mathrm{d}\bm{r}' n_\mu(\bm{r}', \beta)\ln{\left(\frac{n_\mu(\bm{r}', \beta)}{N_\mu}\right)}\,. \label{ent_funcs}
\end{align}
We shall examine the intuitive polymeric interpretation of spherical symmetry-breaking in the Results  \ref{3} and Discussion  \ref{4} sections.

To solve the set of self-consistent equations \ref{diffuse},\ref{1part}, and \ref{field_dens}, the biggest obstacle is the solution to the modified diffusion equation eqn. \ref{diffuse}, which needs to be solved a number of times corresponding to the total number of pairs $N_p$ in the system --- per self-consistent iteration --- for every value of both spatial positions $\bm{r}$ and $\bm{r}'$. This double spatial dependence of $q_\mu(\bm{r}, \bm{r}', s)$ means that traditional real-space methods are impractical for computational efficiency \cite{thompson2019.article,Matsen_Schick1994.article,matsen.incollection}. Instead, what is usually done in the polymer SCFT community, is to use the spectral method: all spatially-dependent functions are decomposed in terms of a set of basis functions, for which the position dependence of each function can be integrated out and the resulting equations become matrix equations for the unknown expansion coefficients. The problem encountered earlier with real-space methods is then made to vanish and is replaced with solving a matrix equation $N_p$ times per self-consistent iteration \cite{Matsen_Schick1994.article,thompson2019.article}. In this work we use the spectral method with non-orthogonal Gaussian basis sets outlined in reference \onlinecite{LeMaitre_Thompson_2022} to solve the modified diffusion equation. The method, along with the numerical procedure, has been discussed in reference \onlinecite{LeMaitre_Thompson_2022}, however, only the case of spherical Gaussian basis functions was treated. To solve the angular problem, we will need the full angular Gaussian basis functions, which are given by the (normalized) expression
\begin{align}
f_{i} (\bm{r}) = \left(2(2c_{pl})^{l+\frac{3}{2}}\left[\Gamma\left(l+\frac{3}{2}\right)\right]^{-1}\right)^{\frac{1}{2}}Z^m_{l} (\theta,\phi) r^l e^{- c_{pl} r^2 }\,, \label{gauss}
\end{align}
where the index $i$ represents the tuple of indices $(p, l, m)$ and $Z^l_{m} (\theta,\phi)$ are the real spherical harmonics defined in appendix \ref{AppendixA}. From here, the basis-specific quantities outlined in appendix \ref{AppendixA} of the overlap matrix (eqn. \ref{over_gauss}), Laplace matrix (eqn. \ref{new_lap_gauss}), and Gamma tensor (eqn. \ref{gamma_gauss}) must be recalculated, but the general non-orthogonal spectral equations listed in appendix \ref{AppendixB} remain unchanged. Although the differences from the spherical case only appear in the basis-specific quantities, it should be pointed out that the addition of angularity increases the dimensions of the basis-specific quantities substantially, severely increasing the computational runtime for even a modest sized basis set, and also introduces a complicated quantity into the expression for the Gamma tensor known as the real Gaunt coefficients; all details can be found in appendix \ref{AppendixA} and references \onlinecite{LeMaitre_Thompson_2022,thompson2019.article}.

The Gaussian exponents $c_{pl}$ are chosen according to an “even-tempered”-type scheme outlined in reference \onlinecite{LeMaitre_Thompson_2022}. In this work $N_b=425$ Gaussian basis functions were used for the angular results since this number provided excellent resolution and converged far enough to the infinite basis set limit, although more basis functions are used in this case compared with reference \onlinecite{LeMaitre_Thompson_2022} because we assigned numbers for each $l$ value of the real spherical harmonics, which in turn have $2l+1$ types of basis functions (for the number of $m$ values associated to each $l$). Convergence here is judged according to the spectral convergence criterion used in reference \onlinecite{LeMaitre_Thompson_2022}. The number 425 comes from assigning 150 basis functions to $l=0$, 50$\times$3 basis functions to $l=1$, and 25$\times$5 basis functions to $l=2$. Increasing $l$ values are assigned smaller numbers of basis functions because the corresponding basis functions become much more diffuse and start to represent smaller portions of the electron density profile. The angular results are also only presented for the first 10 elements, so 150 basis functions for the $l=0$ portion is more than enough to achieve good resolution; the approximations used in this work also limit our accuracy a lot more than the basis set truncation error does. The set of minimum exponents we chose was $c_{1,1,1}=(10^{-15}, 10^{-10}, 10^{-6})$ and the set of maximum exponents was $c_{150,50,25}=(10^{11}, 10^{5}, 10^{3})$, where each entry corresponds to $l$ values in increasing order, respectively. Anything higher than 425 would only add a small amount of resolution and would require a relatively large increase in computation time. For 425 basis functions, every element was able to satisfy a tolerance of at least $10^{-6}$, with some going as far as $10^{-8}$. We decided to keep the same value for $g_0$ that was used in reference \onlinecite{LeMaitre_Thompson_2022} of $g_0=0.1$ for the arbitrary constant $g_0$ associated to the Pauli-exclusion field eqn. \ref{wpauli}, since this will make comparison with previous results easier. The full angular Gaussian basis set used in this work does not introduce or change any previously encountered numerical considerations addressed in reference \onlinecite{LeMaitre_Thompson_2022}. 

\section{Results}
\label{3} 

\begin{figure}
\centering
\includegraphics[height=8cm, width=16cm]{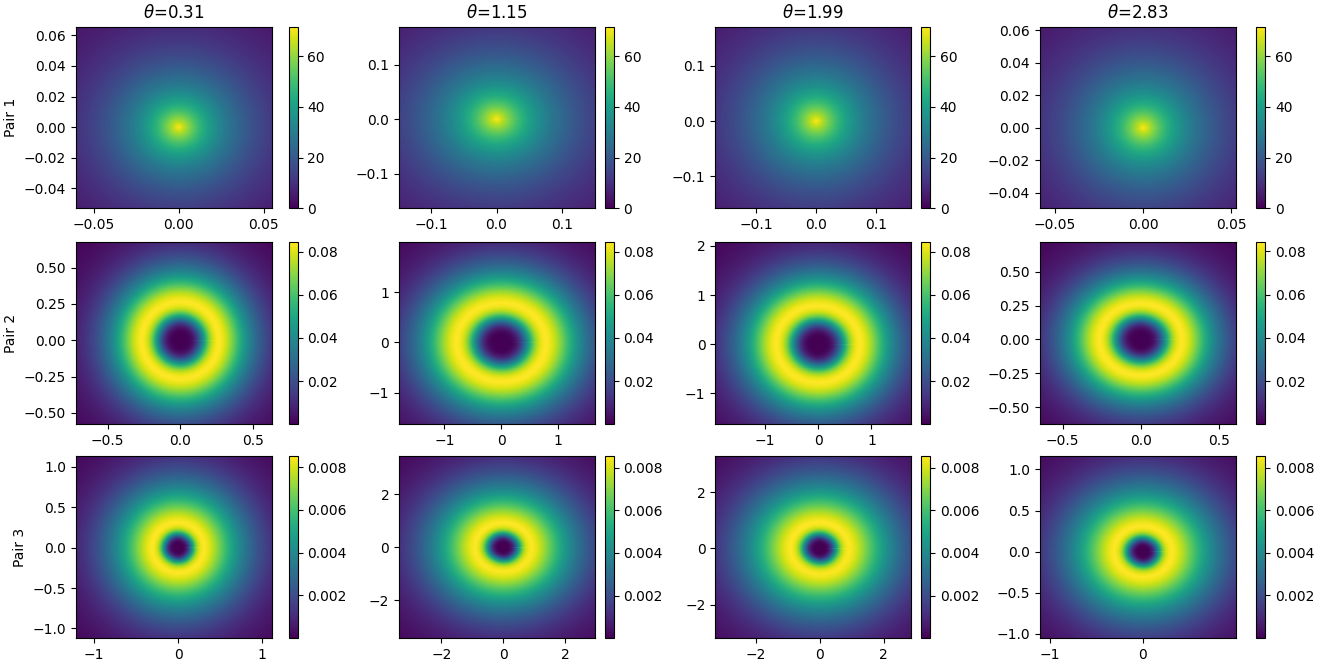}
\caption{Boron angular electron pair density contour plots for fixed values of $\theta$. The axes correspond to $x$ and $y$ Cartesian coordinates and the colour bar indicates the magnitude of the density.}
\label{ang_boron}
\end{figure}

\begin{figure}
\centering
\includegraphics[height=8cm, width=16cm]{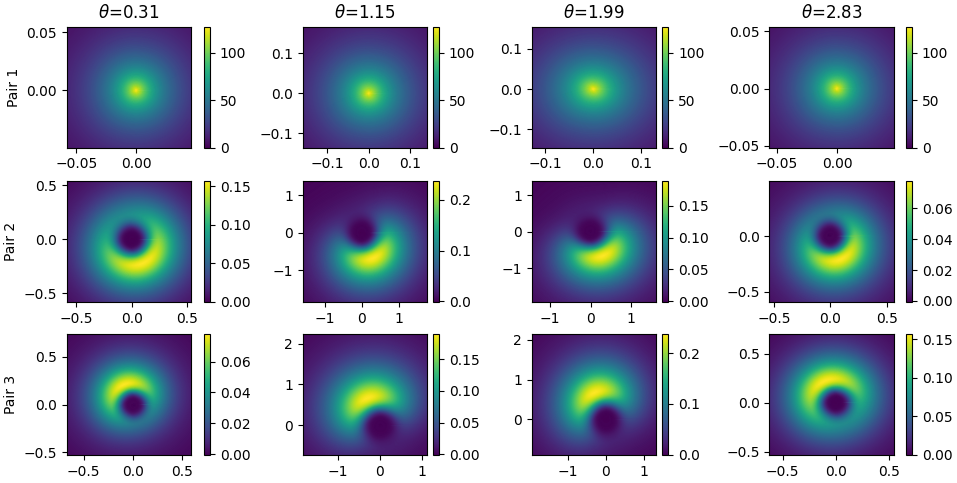}
\caption{Carbon angular electron pair density contour plots for fixed values of $\theta$. The axes correspond to $x$ and $y$ Cartesian coordinates and the colour bar indicates the magnitude of the density.}
\label{ang_carbon}
\end{figure}

\begin{figure}
\centering
\includegraphics[height=11cm, width=16cm]{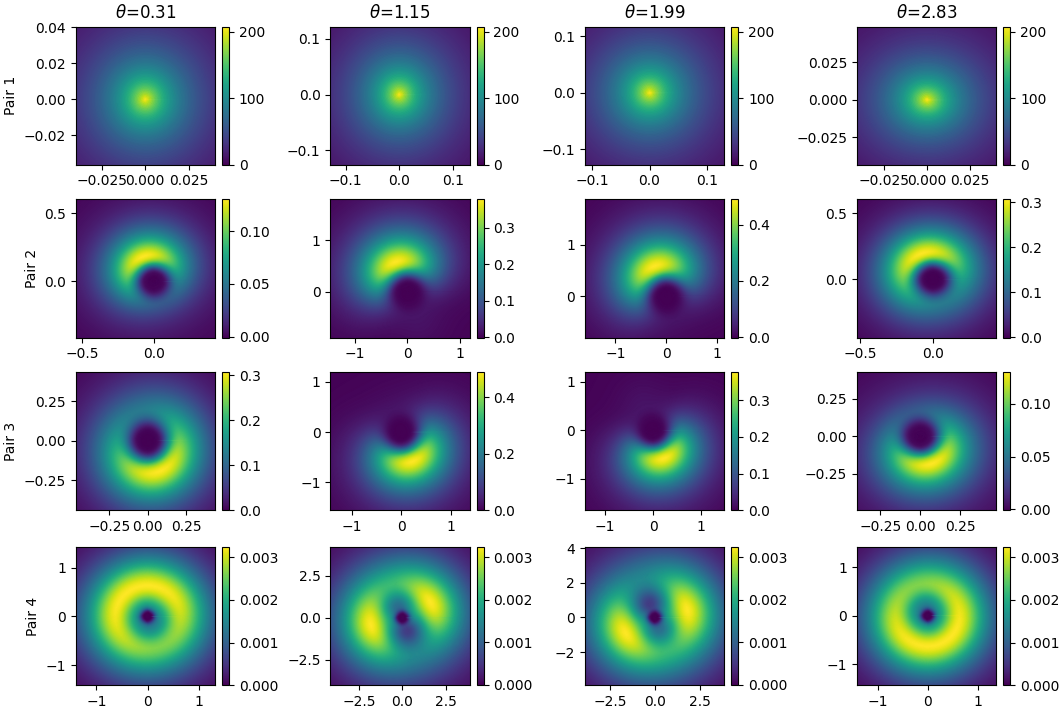}
\caption{Nitrogen angular electron pair density contour plots for fixed values of $\theta$. The axes correspond to $x$ and $y$ Cartesian coordinates and the colour bar indicates the magnitude of the density.}
\label{ang_nitrogen}
\end{figure}

\begin{figure}
\centering
\includegraphics[height=11cm, width=16cm]{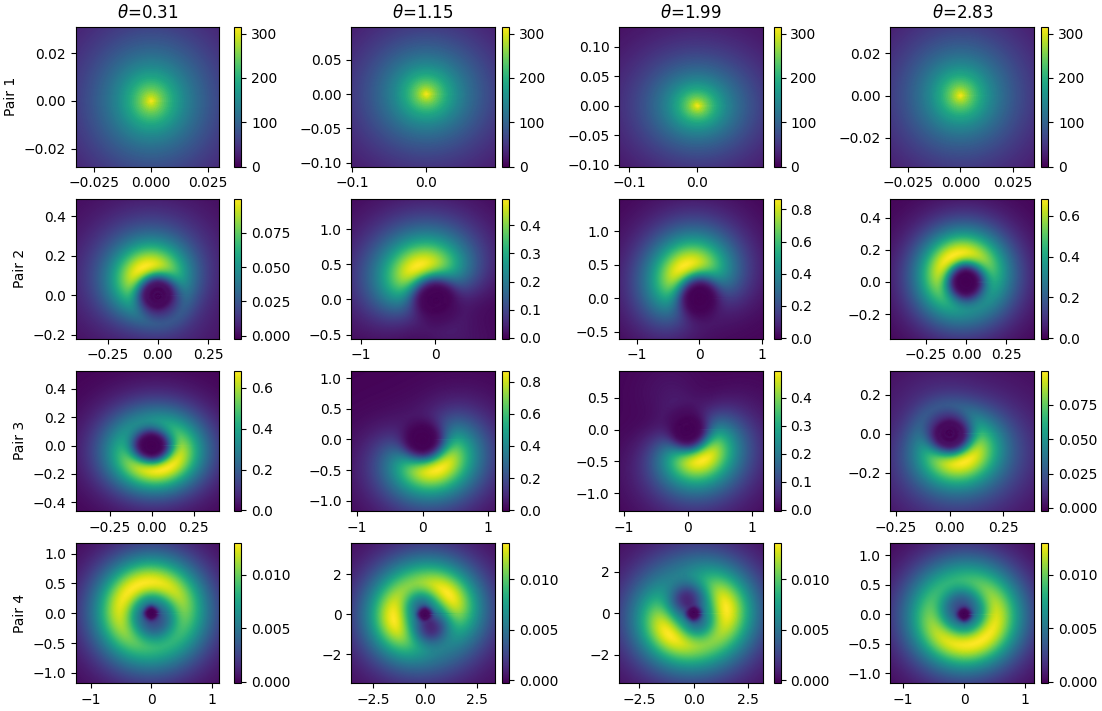}
\caption{Oxygen angular electron pair density contour plots for fixed values of $\theta$. The axes correspond to $x$ and $y$ Cartesian coordinates and the colour bar indicates the magnitude of the density.}
\label{ang_oxygen}
\end{figure}

\begin{figure}
\centering
\includegraphics[height=13cm, width=16.2cm]{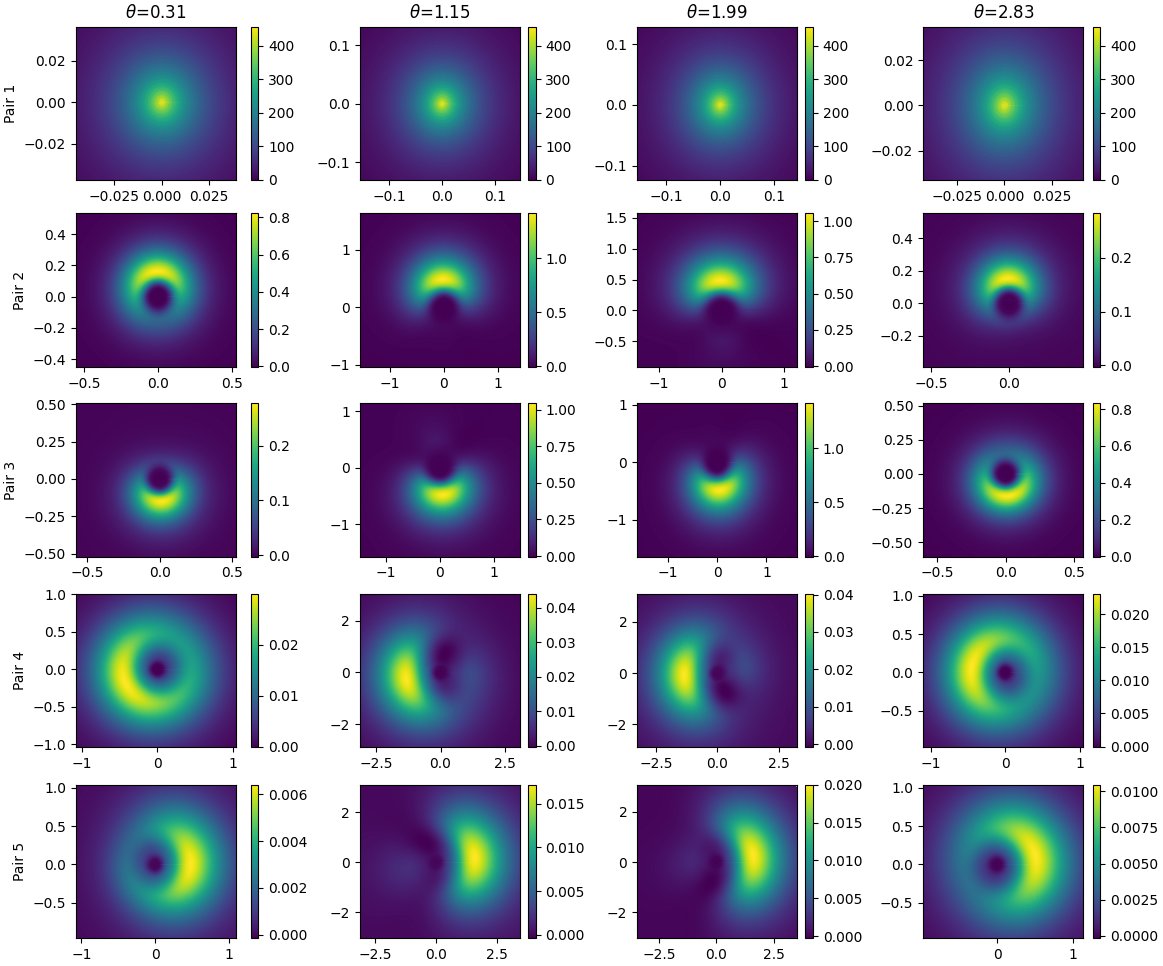}
\caption{Fluorine angular electron pair density contour plots for fixed values of $\theta$. The axes correspond to $x$ and $y$ Cartesian coordinates and the colour bar indicates the magnitude of the density.}
\label{ang_fluorine}
\end{figure}

\begin{figure}
\centering
\includegraphics[height=13cm, width=16.2cm]{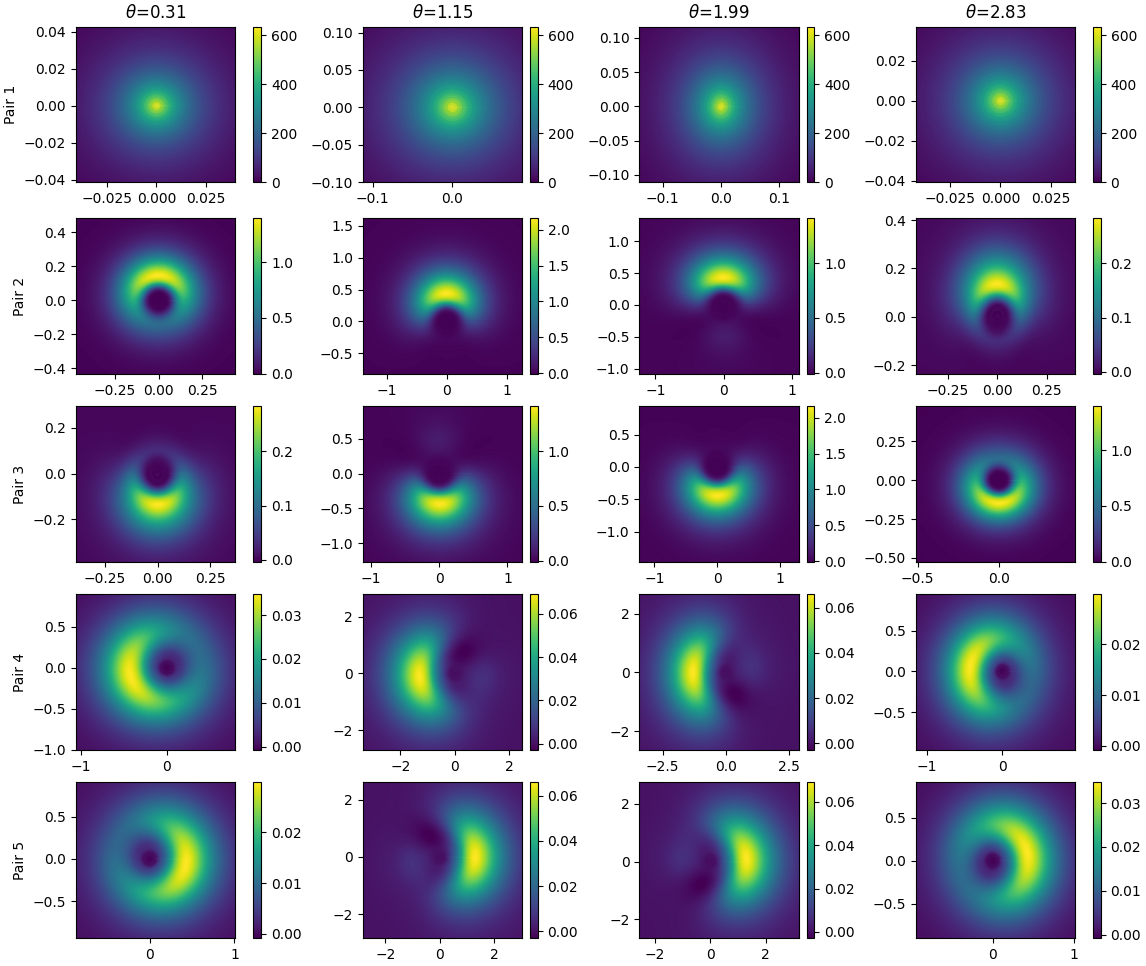}
\caption{Neon angular electron pair density contour plots for fixed values of $\theta$. The axes correspond to $x$ and $y$ Cartesian coordinates and the colour bar indicates the magnitude of the density.}
\label{ang_neon}
\end{figure}

The atomic binding energies corresponding to $g_0=0.1$ for the elements hydrogen to neon, strictly enforcing the maximum occupancy of 2 electrons per pair, are shown in table \ref{tab2} for the spectral expansion of the density with angular basis functions, and table \ref{tab3} for the restriction to spherically-symmetric basis functions; atomic units are used unless otherwise specified. The binding energies predicted by our model are contrasted with those predicted by Hartree-Fock theory, since in our neglect of correlations and use of an exact self-interaction correction, Hartree-Fock theory should be considered as ``exact''. One difference with the results from reference \onlinecite{LeMaitre_Thompson_2022} can be seen by looking at the percentage deviation with Hartree-Fock for the two tables: The model with angular dependence is much closer to Hartree-Fock for the first 6 elements, but rapidly worsens due to an overabundance of Pauli-exclusion repulsion felt between electron pairs. In the angular case considered here, the atomic shell configurations are not prebuilt into the code, so the overabundance of the Pauli-exclusion force causes the pairs after carbon to spread too far apart. However, the observed shell structure does arise solely from the information provided by the electron pair configurations and still somewhat resembles what we expect (figures \ref{ang_boron}-\ref{ang_neon})\footnote{The density contour plots of the first four elements are not shown as they are all spherical and have identical pair to shell structure. Reference \onlinecite{LeMaitre_Thompson_2022} can be consulted for plots of hydrogen to beryllium using the model from this work.}. Moreover, we do see spontaneous spherical symmetry-breaking, which is first predicted to occur at carbon as opposed to boron. The shapes of the pair densities after boron do not match naive expectations; this is addressed later in this section. The magnitude of the symmetry-breaking effect can also be modified through the value of $g_0$ (i.e. a smaller value produces more noticeable deviations). In particular, if $g_0$ is given pair dependence, then a simple arithmetic sequence of increasing values chosen from the neighbourhood around $0.1$ such that the smallest value is assigned to the first pair, is sufficient for boron to break spherical symmetry. Although nature predicts spherical symmetry-breaking to first occur at boron, this does not mean the polymer excluded-volume picture is invalid: symmetry-breaking has a subtle effect on the binding energies, as can be seen by comparing tables \ref{tab3} and \ref{tab2}, so the approximation used for the Pauli-exclusion field may be too coarse to allow such a prediction, in which case the exact field would need to be implemented to sufficiently test this. The fact that spontaneous spherical symmetry-breaking does occur, and only one element off from where it is supposed to be, is a very encouraging result.
\begin{longtable}{ p{.09\textwidth}  p{.12\textwidth}  p{.15\textwidth}  p{.14\textwidth} }
\hline
Element & SCFT & Hartree-Fock & \% Deviation \\
\hline
H & 0.4999999 & 0.5000000000 & 0.000002\\
He & 2.861679 & 2.861679996 & 0.000000014\\
Li & 7.46842 & 7.432726931 & 0.47792\\
Be & 14.70219 & 14.57302317 & 0.87856\\
B & 24.66954 & 24.52906073 & 0.56944\\
C & 37.655254 & 37.68861896 & 0.088607\\
N & 53.65814 & 54.40093421 & 1.38431\\
O & 72.8257 & 74.80939847 & 2.7239\\
F & 95.2256 & 99.40934939 & 4.3935\\
Ne & 120.9975 & 128.5470981 & 6.2394\\
\hline
\caption{Atomic binding energies for hydrogen to neon using full angular basis sets with pair occupancy restricted to a maximum of 2. The number of decimal places for the binding energies predicted by our model correspond to the numerical accuracy of our calculations, where the numerical uncertainty is expressed in the last digit. The Hartree-Fock binding energies are taken from \cite{koga_thakker.article}.} \label{tab2} 
\end{longtable}
\begin{longtable}{ p{.09\textwidth}  p{.12\textwidth}  p{.15\textwidth}  p{.14\textwidth} }
\hline
Element & SCFT & Hartree-Fock & \% Deviation \\
\hline
H & 0.4999999 & 0.5000000000 & 0.000002\\
He & 2.861679 & 2.861679996 & 0.000000014\\
Li & 7.46842 & 7.432726931 & 0.47792\\
Be & 14.70219 & 14.57302317 & 0.87856\\
B & 24.66954 & 24.52906073 & 0.56944\\
C & 37.567740 & 37.68861896 & 0.321764\\
N & 53.40706 & 54.40093421 & 1.86094\\
O & 72.3335 & 74.80939847 & 3.4229\\
F & 94.3264 & 99.40934939 & 5.3887\\
Ne & 119.5084 & 128.5470981 & 7.5633\\
\hline
\caption{Atomic binding energies for hydrogen to neon using spherically-symmetric basis sets with pair occupancy restricted to a maximum of 2. The number of decimal places for the binding energies predicted by our model correspond to the numerical accuracy of our calculations, where the numerical uncertainty is expressed in the last digit. The Hartree-Fock binding energies are taken from \cite{koga_thakker.article}.} \label{tab3} 
\end{longtable}
There is no difference in the binding energies between the spherical and angular results for the first 4 elements since these elements are known to have spherical ground-state distributions and minimal Pauli-exclusion repulsion. The lack of a binding energy difference between the spherical and non-angular cases of boron is attributed to the approximate Pauli potential used in this work, which predicts symmetry-breaking to occur at carbon instead of boron. Carbon is the first element where any difference in the binding energy between the spherical and non-angular cases can be seen: the percent difference of the SCFT model with the prediction from Hartree-Fock theory in the angular case is approximately an order of magnitude smaller than in the spherical case. The agreement for carbon is due in part to the cancellation of certain errors as opposed to a genuine agreement with Hartree-Fock theory, but the other trends in tables \ref{tab3} and \ref{tab2} suggest that the angular case does yield an improvement in the agreement with Hartree-Fock theory. The rest of the elements from tables \ref{tab3} and \ref{tab2} display very minor changes in the binding energy, which agrees nicely with the results of Chowdhury and Perdew \cite{Chowdhury_Perdew2021}, who report that the effect of symmetry-breaking has a small impact on the binding energy. These results suggest that the spherical-averaging approximation used in reference \onlinecite{LeMaitre_Thompson_2022} performs quite well in most scenarios, and that it is physically reasonable to use it for isolated atoms provided the aim is not to investigate delicate features that arise from angularity.

In figures \ref{ang_boron}-\ref{ang_neon}, we see that the first pair density always remains spherically-symmetric, which makes sense as this pair corresponds to the innermost electrons in the atom whose density profile is dominated by the spherical 1s contribution. In fact, because the magnitude of the density for the first pair is so much larger than the rest, the total density profile only marginally deviates from a spherical distribution. The pair densities beyond the first pair do not correspond to the orbital picture we get from other DFT approaches, and the non-spherical pair densities only resemble a single lobe in contrast to the multiple lobes expected from orbital pictures such as Hartree-Fock theory. However, pair densities do not correspond to the squared modulus of individual orbitals from KS-DFT, rather, they correspond to sums of squared moduli of orbitals, as shown in appendix \ref{AppendixC}. The pair density profiles corresponding to non-spherical pairs somewhat resemble the situation in polymer macro-phase separation \cite{matsen.incollection}, where one pair occupies one of the lobes in one region and the other occupies the partner lobe across from it, together forming a hybrid 2s-2p-like structure. This macro-phase behaviour is unsurprising given that the system is being modelled as a classical polymeric system with higher-dimensional excluded-volume, a model which is equivalent to the wavefunction picture through the theorems of DFT \cite{Thompson2022}. The total densities of the atoms seen in figures \ref{ang_carbon}-\ref{ang_neon} have approximately the same profiles as the densities predicted by quantum mechanics in the wavefunction picture, but due to the inexact Pauli-exclusion field used in this work, we do not expect to produce identical profiles.

In light of the macro-phase-like behaviour seen in figures \ref{ang_carbon}-\ref{ang_neon}, there are a number of constraints that the true electron density must satisfy in order to be considered physically acceptable, which we can use as benchmarks to assess the density profiles predicted by the model used in this work. The two main constraints on the electron density are that it must be positive for all positions $\bm{r}$ and that its integral over all space must yield the electron number $N$ (or the pair electron number if we are dealing with an individual pair density). Two further constraints are 
\begin{align}
1 \geq \frac{3\pi}{4K}\left[\frac{\pi}{2}\int \mathrm{d}\bm{r}\, n^3(\bm{r}, \beta)\right]^{\frac{1}{3}}\ \ \text{and} \ \ 1 \geq \frac{1}{2K}\int \mathrm{d}\bm{r}\, \left(\nabla \sqrt{n(\bm{r}, \beta)}\right)^2 \label{den_constr}
\end{align}   
where $K$ represents the kinetic energy of the system \cite{lieb2002}, which is given by eqn. \ref{kin_en} in the model. The first constraint of eqns. \ref{den_constr} is the requirement that the electron density be contained in the function space $L^3$, while the second is the requirement that the kinetic energy associated with the electron density be bounded below by the von Weizsacker kinetic energy \cite{lieb2002}. 

\begin{longtable}{ p{.09\textwidth}  p{.13\textwidth}  p{.13\textwidth} }
\hline
Element & Constraint 1 & Constraint 2 \\
\hline
H & 0.85127 & 0.99985 \\
He & 0.87446 & 0.99983 \\
Li & 0.85268 & 0.95681 \\
Be & 0.83296 & 0.92839 \\
B & 0.82156 & 0.91523 \\
C & 0.80450 & 0.90587 \\
N & 0.79150 & 0.89451 \\
O & 0.78058 & 0.88589 \\
F & 0.77055 & 0.87868 \\
Ne & 0.76157 & 0.87234 \\
\hline
\caption{The values for the right-hand side of eqns. \ref{den_constr} for hydrogen to neon using full angular basis sets and populating the pairs according to their original definition.} \label{tab5} 
\end{longtable}

The pair density profiles corresponding to figures \ref{ang_boron}-\ref{ang_neon} clearly demonstrate that the pair densities, and thus the total density, are non-negative for all positions $\bm{r}$, since we know the density goes to zero for large $r$. Likewise, the expected electron numbers corresponding to each pair were obtained from the pair density integrals to within numerical accuracy (e.g. basis set truncation), adding up to the desired total electron number in every case. Table \ref{tab5} displays the right-hand side values of eqns. \ref{den_constr} for the elements hydrogen to neon, where we can see that the density predicted by the model always satisfies these two inequalities. In particular, one can notice that the right-hand side values of the second inequality in eqns. \ref{den_constr} for hydrogen and helium are almost exactly 1, which is the statement that the von Weizsacker kinetic energy functional is exact for one and two electron systems.

\begin{longtable}{ p{.07\textwidth} p{.05\textwidth}  p{.11\textwidth}  p{.1\textwidth}  p{.1\textwidth}  p{.09\textwidth}  p{.11\textwidth}  p{.1\textwidth}  p{.09\textwidth} p{.1\textwidth} }
\hline
 & & $U_{e-n}$ & $U_{e-e}$ & $U_{\text{sic}}$ & $U_{P}$ & $U$ & $-S_{c}/\beta$ & $-S_{t}/\beta$ & $F$ \\
\hline
\multirow{2}{*}{Pair 1} & Sph. & -69.75559 & 9.91254 & -3.62593 & 0.31151 & -63.15747 & 33.78870 & 0.00376 & -29.36501 \\
 & Ang. & -69.70639 & 10.05499 & -3.62318 & 0.29381 &  -62.98077 & 33.74174 & 0.00373 & -29.23530 \\
\hline
\multirow{2}{*}{Pair 2} & Sph. & -10.06770 & 3.81522 & -0.63213 & 0.48931 & -6.39530 & 1.29296 & -0.01316 & -5.11551 \\
 & Ang. & -8.43783 & 3.47576 & -0.58622 &  0.32278 & -5.22551 & 1.02975 & -0.01423 & -4.20999 \\
\hline
\multirow{2}{*}{Pair 3} & Sph. & -5.92133 & 2.56803 & -0.35264 & 0.27463 & -3.43131 & 0.36323 & -0.01914 & -3.08723 \\
 & Ang. & -8.43782 & 3.47575 & -0.58622 & 0.32278 & -5.22551 & 1.02975 & -0.01423 & -4.20998 \\
\hline
\multirow{2}{*}{Total} & Sph. & -85.74463 & 16.29579 & -4.61070 & 1.07545 & -72.98408 & 35.44489 & -0.02854 & -37.56774 \\
 & Ang. & -86.58204 & 17.00649 & -4.79562 & 0.93938 & -73.43179 & 35.80124 & -0.02473 & -37.65527 \\
\hline
\caption{Free energy, potential energy, and the entropic contributions to the free energy corresponding to each pair in the element carbon.} \label{tab4} 
\end{longtable}

\begin{longtable}{ p{.07\textwidth} p{.05\textwidth}  p{.11\textwidth}  p{.1\textwidth}  p{.1\textwidth}  p{.09\textwidth}  p{.11\textwidth}  p{.1\textwidth}  p{.09\textwidth} p{.11\textwidth} }
\hline
 & & $U_{e-n}$ & $U_{e-e}$ & $U_{\text{sic}}$ & $U_{P}$ & $U$ & $-S_{c}/\beta$ & $-S_{t}/\beta$ & $F$ \\
\hline
\multirow{2}{*}{Pair 1} & Sph. & -161.60445 & 16.80288 & -5.63552 & 0.91353 & -149.52355 & 80.61756 & 0.00821 & -68.89778 \\
 & Ang. & -161.7626 & 17.26576 & -5.64371 & 0.91018 & -149.23038 & 80.79204 & 0.0081 & -68.43023 \\
\hline
\multirow{2}{*}{Pair 2} & Sph. & -24.96449 & 7.58292 & -1.08034 & 1.72185 & -16.74006 & 4.12557 & -0.00759 & -12.62207 \\
 & Ang. & -21.39702 & 7.25637 & -1.11264 & 1.0224 & -14.23088 & 3.69921 & -0.00781 & -10.53948 \\
\hline
\multirow{2}{*}{Pair 3} & Sph. & -10.73121 & 4.40419 & -0.43666 & 0.71924 & -6.04444 & 0.53809 & -0.01693 & -5.52329 \\
 & Ang. & -21.39702 & 7.25638 & -1.11264 & 1.0224 & -14.23088 & 3.69922 & -0.00781 & -10.53948 \\
\hline
\multirow{2}{*}{Pair 4} & Sph. & -10.73121 & 4.40419 & -0.43666 & 0.71924 & -6.04444 & 0.53809 & -0.01693 & -5.52329 \\
 & Ang. & -7.67115 & 3.47493 & -0.35324 & 0.21603 & -4.33343 & 0.3893 & -0.01902 & -3.96315 \\
\hline
\multirow{2}{*}{Pair 5} & Sph. & -3.40092 & 1.53289 & -0.13416 & 0.15004 & -1.85215 & 0.10301 & -0.01092 & -1.76006 \\
 & Ang. & -3.38691 & 1.55932 & -0.17473 & 0.08222 & -1.9201 & 0.17628 & -0.00962 & -1.75344 \\
\hline
\multirow{2}{*}{Total} & Sph. & -211.43228 & 34.72707 & -7.72335 & 4.22391 & -180.20465 & 85.92232 & -0.04415 & -94.32649 \\
 & Ang. & -215.6147 & 36.81275 & -8.39695 & 3.25323 & -183.94568 & 88.75605 & -0.03616 & -95.22579\\
\hline
\caption{Free energy, potential energy, and the entropic contributions to the free energy corresponding to each pair in the element fluorine.} \label{tab6} 
\end{longtable}

Tables \ref{tab4} and \ref{tab6} list the pairwise components of the potential energy terms, the energy contributions from the configurational and translational entropies, and the total free energies for the elements carbon and fluorine, respectively. In both elements we can see that the lowering of the free energy due to spherical symmetry-breaking is produced from pairs 2 and 3, which both adopt opposing lobe shapes, in contrast to the purely spherical distributions in the spherically-averaged case, that distributes their free energy contribution uniformly amongst the two. It is this feature in particular that accounts for the free energy difference between the spherical and angular cases. Looking more closely at tables \ref{tab4} and \ref{tab6}, the cause of this feature is the fact that the third pair density can occupy a region closer to the atomic nucleus by violating spherical symmetry, which is evidenced by the much lower $U_{e-n}$ value in both elements for the angular case. The electron-electron plus self-interaction correction potential, the Pauli potential, and both entropic contributions to the free energy for pair 3 are all worse in the angular case, suggesting that the move towards the nucleus more than compensates for the interaction penalties with other pairs. In order for pair 3 to accomplish the move from a spherical distribution to a lobe distribution, pair 2 must also transform to a mirroring lobe distribution so that its original spherical shape does not overlap as much with the new lobe shape of pair 3; this effect is propagated with the other pairs (except the first), converting them into non-spherical distributions as well.  

\section{Discussion}
\label{4} 

Despite the failure of the Pauli-exclusion field to satisfy the coordinate scaling relation eqn. \ref{pauli_cons}, the scaling argument presented in reference \onlinecite{LeMaitre_Thompson_2022} shows that the picture of higher-dimensional excluded-volume interactions between pairs of threads recovers the Thomas-Fermi quantum kinetic energy term and the Dirac exchange term in the uniform density limit. The analysis from section \ref{3} also demonstrates that the pair densities making up the total density satisfy all constraints necessary to guarantee a physically acceptable electron density. Together with the proof that pair densities do not necessarily correspond to individual squared moduli of orbitals, and the formal equivalence of the polymer-thread picture with quantum DFT through the quantum-classical isomorphism \cite{Feynman1953AtomicTO}, the macro-phase-like behaviour exhibited in figures \ref{ang_carbon}-\ref{ang_neon} show that spontaneous shell structure and spherical-symmetry breaking are robust predictions. The symmetry-breaking arises from the energetic benefit of electrons distributing asymmetrically closer to the nucleus. It is also clear from adding up the individual pair density profiles for any given atom, that the total density profiles only deviate slightly from spherical symmetry, which is consistent with the findings of Chowdhury and Perdew \cite{Chowdhury_Perdew2021} that asymmetries in the electron density have a small effect. Together, these two observations highlight an important distinction: the ability of the pairs to individually break spherical-symmetry allows the atom to lower its binding energy in all cases, but this does not necessarily mean that the total electron density also breaks spherical symmetry. 

One should consider whether the macro-phase-like behaviour is simply an artefact of the specific approximation for the Pauli potential being used in this work, as it is reasonable to speculate that the exaggerated repulsion produced by the approximate Pauli-exclusion field causes the pairs to clear their local neighbourhood. That is, are we only observing isolated atoms to be spherically-asymmetric because we are using an approximate model? If we implemented the exact Pauli potential, would total electron densities always be found to be spherically symmetric, consistent with the arguments of references \onlinecite{Cohen1965, gil1972}?. This seems unlikely, since this would require an unjustifiable perfect balance between the Pauli potential and other factors. Other frustrated systems induce spontaneous symmetry breaking, for example in true polymeric systems, SCFT is used to predict the micro-phases of block copolymers \cite{Matsen_Schick1994.article, Matsen_Bates1997, matsen.incollection}. Returning to the present model, the shell structure for carbon displayed in figure \ref{ang_carbon} demonstrates non-spherical structure yet maintains nearly identical shell structure to that predicted by Hartree-Fock theory. As was mentioned earlier, there is probably some cancellation of errors happening within carbon due in part to competing Pauli pair repulsions, but it seems unlikely based on the trends from the other atoms combined with the difference in sensitivity between the binding energies and the density profiles, that the magnitude of this effect would be large enough to account for the macro-phase-like structure while also producing the minute differences with Hartree-Fock theory that are observed. 

\section{Conclusions and Future Work}
\label{5}   

The ring polymer SCFT formulation of quantum mechanics predicts the spontaneous emergence of atomic shell structure and spherical symmetry-breaking in isolated neutral atoms hydrogen to neon Using postulated pair structure of the model and ideas of higher-dimensional excluded-volume in cooperation with an exact self-interaction correction, the model shows excellent agreement with Hartree-Fock theory for the atomic binding energies and density profiles of the first six elements, providing exact matches for the elements hydrogen and helium. However, due to the approximation made on the Pauli-exclusion field, the predicted shell structure starts to deviate significantly past the element neon and the symmetry-breaking is first predicted to occur at carbon instead of boron. Consistent with Chowdhury and Perdew \cite{Chowdhury_Perdew2021}, the symmetry-breaking effect is found to have a very small impact on the binding energies, which suggests that the spherical-averaging approximation is physically reasonable when investigating atomic systems. The pair density contour plots also display behaviour similar to polymer macro-phase separation, where individual electron pairs occupy single lobe structures that together form a dumbbell shape analogous to the 2p orbital shape. It is further shown that the predicted densities satisfy known constraints and still produce the same total electronic density profile that is predicted by quantum mechanics.  

There are a number of future directions to consider, now that the basic engine from reference \onlinecite{LeMaitre_Thompson_2022} has been constructed. One possible direction could be to extend the work of reference \onlinecite{sillaste_thompson.article} in modelling systems of diatomic molecules to arbitrary formations of molecules or even solid-state lattices, since the Gaussian methodology is easily adapted to any number of complex geometries. The initial groundwork involved in this direction would involve switching to contracted Gaussian basis sets \cite{Helgaker_Taylor1995,huzinaga2012gaussian,Hehre_Pople_Stewart1969}, since they allow for many fewer basis functions to be used while still maintaining roughly the same level of precision and resolution; the price tag associated to the contracted sets comes in the form of additional minimization routines that either minimize the spectral representation of the free energy eqn. \ref{free_en} with respect to even-tempered parameters \cite{Schmidt_Ruedenberg1979,Helgaker_Taylor1995} or fit the Gaussians to a Slater-type function \cite{Hehre_Pople_Stewart1969,Helgaker_Taylor1995}. Both of the these methods typically require derivative information to perform the minimization, which is undesirable because the derivatives may not be well-defined or could possibly lead to numerical instabilities yielding false minima. A method that uses only the Nelder-Mead algorithm was originally developed for this work in anticipation of investigating molecular systems, which uses the spectral coefficients of the density and the Gaussian exponents from the uncontracted result to solve for the contraction coefficients, and then minimizes the sum of squared deviations between the two to solve for the exponents of the contracted set. After implementing the contracted sets and updating the current computational engine, one would need to generalize the centres of the Gaussian basis sets to arbitrary positions and modify the structure of the computation to accommodate multiple atoms. The shifting of the centres of the Gaussians from the origin to arbitrary positions is not so easily done with the spherical harmonic representation used in this work and might be better facilitated using a Cartesian representation of the Gaussians \cite{Helgaker_Taylor1995}, which only entails re-deriving the basis specific matrices and the spectral components of the electron-nucleus field. Fortunately, the expressions for these quantities have already been derived, although the expressions are much more complicated. In the case of solid-state lattices, pseudo-potentials and other modifications would need to be introduced as well \cite{Karasiev_Trickey2012,witt_del_rio_dieterich_carter2018}. Further increases in accuracy could also be achieved by combining the molecular dynamics framework of Car and Parrinello \onlinecite{car1985unified.article} with the model, to better account for the nuclear degrees of freedom. 

Another possible direction is to implement an exact Pauli-exclusion field, so that the comparison of the present model with Hartree-Fock theory can be completed, and the equivalence of higher-dimensional excluded-volume with the Pauli-exclusion principle can be verified. This direction would be an important test of a foundational aspect of the theory \cite{Thompson2022}, and would provide further evidence in support of the symmetry breaking mechanism presented here, in which electrons lower the free energy by breaking spherical symmetry to form electron distributions that approach closer to the atomic nucleus. 

Other possible directions include adding correlation terms and relativistic corrections, or clarifying the mechanism for electron spin in the model. Correlation fields are stipulated to be the only thing missing from this model that prevents it from completely agreeing with the predictions of non-relativistic quantum mechanics for neutral atoms not in the presence of any other fundamental fields. Therefore, finding a mechanism for correlations within the polymer picture would also be of foundational importance to the model. An investigation of quantum correlations would also naturally lead to the topic of quantum entanglement, which further connects with information-theoretic approaches to DFT \cite{Nagy2015,ALIPOUR2015} that may be useful in learning more about properties of correlations in many-body systems from the perspective of DFT. Investigating the mechanism that represents electron spin would also complement the study of correlations. Lastly, relativistic corrections including fine structure \cite{Liu2020,Autschbach2012} and finite-size nuclear centres \cite{Visscher_Dyall} could be added to the model in order to study heavier elements, which may yield useful information on how relativistic effects manifest themselves in the polymer picture.

\section*{Acknowledgements}
This research was financially supported by the Natural Sciences and Engineering Research Council of Canada (NSERC).

\appendix
\section*{APPENDICES}
\section{Spectral Method}
\label{AppendixB} 

Instead of using real-space methods to solve the SCFT equations, which are computationally very inefficient due to the double spatial dependence of the single particle propagator, all spatially-dependent functions can be expanded in terms of a complete set of basis functions $\{f_i(\bm{r})\}^{\infty}_{i=1}$. The choice of basis functions is arbitrary, with some choices better than others for particular problems, and may even be chosen to be non-orthogonal, which is the strategy adopted in reference \onlinecite{LeMaitre_Thompson_2022}. In this section, the spectral expressions for the SCFT equations taken from reference \onlinecite{LeMaitre_Thompson_2022} will merely be stated for convenience; further details can be found in reference \onlinecite{LeMaitre_Thompson_2022}.   

For a general function of one spatial variable $g(\bm{r})$ is expressed as 
\begin{align}
g(\bm{r}) = \sum_i g_i f_i(\bm{r})
\end{align}
where $g_i$ are the spectral expansion coefficients. A general function of two spatial variables $h(\bm{r}, \bm{r}')$ is expressed as
\begin{align}
h(\bm{r}, \bm{r}') = \sum_{ij} h_{ij} f_i(\bm{r})f_j(\bm{r}') \label{biexp}
\end{align}
where $h_{ij}$ are the spectral expansion coefficients. 

Before giving the spectral expressions for the SCFT equations, we first define the components of the three quantities that will appear throughout the derivation; namely the overlap matrix
\begin{align}
\text{S}_{ij} = \int \mathrm{d}\bm{r}\, f_i(\bm{r})f_j(\bm{r}) \label{overlap}
\end{align}
Laplace matrix
\begin{align}
\text{L}_{ij} = \int \mathrm{d}\bm{r}\, f_i(\bm{r})\nabla^2f_j(\bm{r}) \label{laplace}
\end{align}
and the gamma tensor
\begin{align}
\Gamma_{ijk} = \int \mathrm{d}\bm{r}\, f_i(\bm{r})f_j(\bm{r})f_k(\bm{r}) . \label{gamma}
\end{align}
Just as in reference \onlinecite{LeMaitre_Thompson_2022}, we will adopt a matrix notation where spectral coefficients will be referred to in bolded font. Thus, eqns. \ref{overlap}-\ref{gamma} will be referred to using the symbols $\bf{S}, \bf{L},\ \text{and} \ \bm{\Gamma}$. Quantities dependent on electron pairs will be the only things denoted with an index. 

The spectral SCFT equations for the $\mu$th pair are then given as
\begin{align}
\frac{\mathrm{d}\bf{q}_\mu(s)}{\mathrm{d}s} &= \bf{S}^{-1}\bf{A}_\mu\bf{q}_\mu(s) \label{sqp}\\[1.5ex]
\bf{q}_\mu(0) &= \bf{S}^{-1} \\[1.5ex]
Q_\mu[w](\beta) &= \text{Tr}\left(\bf{S}\bf{q}_\mu(\beta)\right) \\[1.5ex]
\bf{n}_\mu(\beta) &= \frac{N_\mu}{Q_\mu[w](\beta)}\bf{S}^{-1}\bm{\Gamma}\bf{q}_\mu(\beta) \\[1.5ex]
\bf{w}_\mu(\beta) &= \bf{w}^{e-n}_\mu(\beta) + \bf{w}^{e-e}_\mu(\beta) + \bf{w}^{\text{sic}}_\mu(\beta) + \bf{w}^{P}_\mu(\beta) \label{sw}
\end{align} 
where $\bf{q}_\mu(s)$ is the matrix of single-particle propagator spectral coefficients, $Q_\mu[w](\beta)$ is the single-particle partition function, $\bf{n}_\mu(\beta)$ is the vector of electron density spectral coefficients, $\bf{w}_\mu(\beta)$ is the vector of field spectral coefficients, and  
\begin{align}
\bf{A}_\mu = \frac{1}{2}\bf{L} - \bf{w}_\mu\bm{\Gamma}\,.
\end{align}
The individual fields comprising $\bf{w}_\mu(\beta)$ are given by
\begin{align}
\bf{w}^{e-n}_\mu(\beta) &= 4\pi N\bf{L}^{-1}\bf{f}(\bm{0}) \\[1.5ex]
\bf{w}^{e-e}_\mu(\beta) &= -4\pi\bf{L}^{-1}\bf{S}\bm{n}(\beta) \\[1.5ex]
\bf{w}^{\text{sic}}_\mu(\beta) &= \frac{4\pi}{N_\mu}\bf{L}^{-1}\bf{S}\bf{n}_\mu(\beta) \\[1.5ex]
\bf{w}^P_\mu(\beta) &= \frac{1}{g_0} \sum_{\substack{\gamma \\ \gamma \neq \mu}}\bf{n}_\gamma(\beta)
\end{align}
where $\bf{n}(\beta)$ is the vector of total electron density spectral coefficients. The single-particle propagator matrix for each pair $\bf{q}_\mu(s)$ is solved from \ref{sqp} through a generalized eigenvalue problem detailed in reference \onlinecite{LeMaitre_Thompson_2022}, to yield
\begin{align}
\bf{q}_\mu(s) = \bf{U}_\mu e^{\bf{D}_\mu s}\bf{U}^T_\mu \label{prop_decomp}
\end{align}
where $\bf{U}_\mu$ is the matrix of generalized eigenvectors and $\bf{D}_\mu$ is the diagonal matrix of generalized eigenvalues. Equation \ref{prop_decomp} then allows for the single-particle partition function $Q_\mu[w](\beta)$ to be rewritten in the more computationally convenient form
\begin{align}
Q_\mu(\beta) = \text{Tr}\left(e^{\bf{D}_\mu \beta}\right)\,. \label{new_part}
\end{align}
Finally, the free energy expression can be calculated as 
\begin{align}
F[n,w] &= -\sum_{\mu}\left[\frac{N_{\mu}}{\beta}\ln\left[\text{Tr}\left(e^{\bf{D}_\mu \beta}\right)\right] + \frac{1}{2}\bf{S}\bf{n}_\mu(\beta)\left(\bf{w}^P_\mu(\beta)+\bf{w}^{\text{sic}}_\mu(\beta)+\bf{w}^{e-e}_\mu(\beta)\right)\right] \label{free_en}
\end{align}
and the total kinetic energy as
\begin{align}
K[n, w] = -\sum_\mu\frac{N}{2Q_\mu[w](\beta)} \bf{L}\bf{q}_\mu(\beta)\,. \label{kin_en}
\end{align}

\section{Basis Function-Specific Quantities}
\label{AppendixA}

In the following derivations we consider single atomic systems centred at the origin of a spherical coordinate system with coordinates $(r, \theta, \phi)$ where $\{0\leq r, 0\leq \theta \leq \pi, 0\leq \phi < 2\pi\}$. For ease of notation, we will take Latin indices to represent the tuple of indices $(p, l, m)$ indicating the basis function expansion coefficient, angular momentum number, and corresponding $m$ values, respectively. Greek indices will represent the Pauli pairs.

The Gaussian basis functions used in this work are given by the expression
\begin{align*}
f_{i} (\bm{r}) = \mathcal{N}_{pl}Z^m_{l} (\theta,\phi) r^l e^{- c_{pl} r^2 }
\end{align*}
where $\mathcal{N}_{pl}$ are the components of the normalization matrix, $c_{pl}$ are the Gaussian basis exponents, and $Z^m_{l} (\theta,\phi)$ are the real spherical harmonics. Before proceeding with the derivations, we will first define the real spherical harmonics and introduce a few important properties. The standard spherical harmonics are defined as
\begin{align}
 Y^m_{l} (\theta,\phi) = \sqrt{\frac{(2l+1)}{4\pi}\frac{(l-m)!}{(l+m)!}}P^m_l(\cos\theta)e^{im\phi}
\end{align}
where $\{0\leq l, -l \leq m \leq l\}$ and $P^m_l(\cos\theta)$ are the associated Legendre polynomials. The spherical harmonics arise as a solution to the angular part of Laplace's equation $\nabla^{2}u(\bm{r}) = 0$ in spherical coordinates and form an orthonormal basis on the sphere $S^2$, meaning they satisfy the relationship
\begin{align}
\int \int \mathrm{d}\theta\, \mathrm{d}\phi \, Y^m_{l} (\theta,\phi)Y^{m'}_{l'} (\theta,\phi) = \delta_{mm'}\delta_{ll'}\,.
\end{align}
They also obey the parity relation $\,Y^m_{l} (\pi-\theta,\pi+\phi) = (-1)^lY^m_{l} (\theta,\phi)\,$ and are related to their complex conjugate through the relation $\,Y^{m\,*}_{l} (\theta,\phi) = (-1)^mY^{-m}_{l} (\theta,\phi)\,$. The real spherical harmonics are then defined in terms of the standard spherical harmonics by the piecewise expression
\begin{align}
Z^m_{l} (\theta,\phi) =
\begin{cases}
\sqrt{2}\,\text{Re}\left(Y_l^m(\theta, \phi)\right) &,\ \ m > 0\\[1.5ex]
Y_l^m(\theta, \phi) &,\ \ m = 0\\[1.5ex]
\sqrt{2}(-1)^{|m|}\,\text{Im}\left(Y_l^{|m|}(\theta, \phi)\right) &,\ \ m < 0
\end{cases}
\end{align}
where $\text{Re}(z)$ and $\text{Im}(z)$ denote the real and imaginary parts of the argument $z$, respectively. The real spherical harmonics are linear combinations of the standard spherical harmonics that satisfy Laplace's equation, therefore they also satisfy it and preserve many of the same properties that the standard spherical harmonics possess (i.e. orthonormality, parity). One difference lies in their image, which only includes the real numbers as opposed to the complex numbers for the standard spherical harmonics \cite{Homeier_Steinborn1996}. This will prove advantageous when it comes to numerical calculations and plotting of the density profiles, since the extra computer memory required to store complex numbers will not be needed and the basis set will be defined in $\mathbb{R}^3$ as opposed to $\mathbb{R}\otimes\mathbb{C}$ \cite{Homeier_Steinborn1996}. The downside comes from the piecewise definition, meaning some extra consideration will be needed to compute the integral of the product of three real spherical harmonics that appears in the gamma tensor, which will be dealt with in subsection \ref{A3.1.1}.

\subsection{Overlap Matrix}

In order to compute the overlap integral eqn. \ref{overlap}, the coefficients of the normalization matrix $\mathcal{N}_{pl}$ for the Gaussian basis functions must first be computed, which means we must compute the integral: 
\begin{align}
S_{ij} &= \mathcal{N}_{pl}\mathcal{N}_{p'l^{'}}\int^{2\pi}_{0}\int^{\pi}_0\int^{\infty}_0 f_{plm}(\bm{r})f_{p'l^{'}m^{'}}(\bm{r})r^2\sin(\theta)\mathrm{d}r\mathrm{d}\theta\mathrm{d}\phi \nonumber \\[1.5ex] 
&= \mathcal{N}_{pl}\mathcal{N}_{p'l^{'}}\delta_{ll^{'}}\delta_{mm^{'}}\frac{1}{2}\left(\frac{1}{c_{\mu il}+c_{\mu^{'}jl^{'}}}\right)^{\frac{1}{2}(l+l^{'}+3)}\Gamma\left(\frac{l+l^{'}+3}{2}\right)
\end{align}
so
\begin{align}
\mathcal{N}_{pl} = \left(2(2c_{pl})^{l+\frac{3}{2}}\left[\Gamma\left(l+\frac{3}{2}\right)\right]^{-1}\right)^{\frac{1}{2}}\,.
\end{align}
The definition of the normalized Gaussian basis functions then becomes
\begin{align}
f_{i}(\bm{r}) = \left(2(2c_{pl})^{l+\frac{3}{2}}\left[\Gamma\left(l+\frac{3}{2}\right)\right]^{-1}\right)^{\frac{1}{2}}Z^m_{l}(\theta,\phi)r^le^{-c_{pl}r^2}\,.
\end{align}
The overlap integral yields
\begin{align}
S_{ij} &= \left[\frac{2(2c_{p'l^{'}})^{l^{'}+\frac{3}{2}}2(2c_{pl})^{l+\frac{3}{2}}}{\Gamma\left(l+\frac{3}{2}\right)\Gamma\left(l^{'}+\frac{3}{2}\right)}\right]^{\frac{1}{2}}\left(\frac{1}{c_{pl}+c_{p'l^{'}}}\right)^{\frac{1}{2}(l+l^{'}+3)}\frac{\delta_{ll^{'}}\delta_{mm^{'}}}{2}\Gamma\left(\frac{l+l^{'}+3}{2}\right)\,. \label{over_gauss}
\end{align}

\subsection{Laplace Matrix}

Computing the components of the Laplace matrix (eqn. \ref{laplace}) in a Gaussian basis involves computing the Laplacian of a Gaussian, which we will first detail before performing the full computation. Since the real spherical harmonics in the definition of $f_i(\bm{r})$ satisfy the angular part of Laplace's equation, the Laplacian applied to $f_{i}(\bm{r})$ yields
\begin{align}
\bm{\nabla}^2 f_{i}(\bm{r}) &= 2c_{pl} f_{i}(\bm{r})\left[2c_{pl}r^2-(2l+3)\right] \,.
\end{align}

The components of the Laplace matrix in a Gaussian basis are then given by
\begin{align}
L_{ij} &= \int \mathrm{d}\bm{r} f_{j}(\bm{r})\bm{\nabla}^2f_{i}(\bm{r}) = \int \mathrm{d}\bm{r} f_{j}(\bm{r})f_{i}(\bm{r})2c_{pl}\left(2c_{pl}r^2 - 2l-3\right) \nonumber \\[1.5ex]
&= \frac{2c_{pl}}{c_{pl}+c_{p'l'}}\left[c_{pl}(l-l')-c_{p'l'}(2l+3)\right]S_{ji}\,. \label{new_lap_gauss}
\end{align}

\subsection{Gamma Tensor}

The components of the Gamma tensor (eqn. \ref{gamma}) in a Gaussian basis are given by
\begin{align}
\Gamma_{ijk} &= \int \mathrm{d}\bm{r} f_{i}(\bm{r})f_{j}(\bm{r})f_{k}(\bm{r}) = \mathcal{N}_{pl}\mathcal{N}_{p'l^{'}}\mathcal{N}_{p''l^{''}}\int_0^{\pi}\int_0^{2\pi} \mathrm{d}\theta \mathrm{d}\phi Z_{l}^m(\theta, \phi)\nonumber \\[1.5ex] 
&\qquad \qquad\times Z_{l^{'}}^{m^{'}}(\theta, \phi)Z_{l^{''}}^{m^{''}}(\theta, \phi)\sin(\theta)\int_0^\infty \mathrm{d}r\, r^{l+l^{'}+l^{''}+2}e^{-(c_{pl}+c_{p'l^{'}}+c_{p''l^{''}})r^2} \nonumber\\[1.5ex]
&= \frac{\alpha^{mm'm''}_{ll^{'}l^{''}}\mathcal{N}_{pl}\mathcal{N}_{p'l^{'}}\mathcal{N}_{p''l^{''}}}{2\left(c_{pl}+c_{p'l^{'}}+c_{p''l^{''}}\right)^{\frac{1}{2}(l+l^{'}+l^{''}+3)}}\Gamma\left(\frac{l+l^{'}+l^{''}+3}{2}\right)\,, \label{gamma_gauss}
\end{align}
where the symbol $\alpha$ represents the integral of three real spherical harmonics. The computation of $\alpha$ is quite involved, so the computation of its components is left to the following section.

\subsubsection{Integral of Three Real Spherical Harmonics}
\label{A3.1.1}

The integral of three spherical harmonics was first worked out by Gaunt \cite{khersonskii1988quantum}, but is expressed most concisely in terms of the Wigner $3-j$ symbols as
\begin{align}
G^{m_1m_2m_3}_{l_1l_2l_3} &\equiv \int \int \mathrm{d}\theta\, \mathrm{d}\phi \, Y^{m_1}_{l_1} (\theta,\phi)Y^{m_2}_{l_2} (\theta,\phi)Y^{m_3}_{l_3} (\theta,\phi) \sin\theta \nonumber \\[1.5ex]
&= \sqrt{\frac{(2l_1+1)(2l_2+1)(2l_3+1)}{4\pi}}\begin{pmatrix} l_1 & l_2 & l_3 \\ 0 & 0 & 0 \end{pmatrix}\begin{pmatrix} l_1 & l_2 & l_3 \\ m_1 & m_2 & m_3 \end{pmatrix}\,.
\end{align}
The Wigner $3-j$ symbols arise naturally when adding angular momentum values in a multi-electron system, possessing the angular momentum selection rules from atomic and molecular physics as part of their mathematical structure. For reference, the selection rules are $-l_i \leq m_i \leq l_i$, $m_1+m_2+m_3=0$, $l_1+l_2+l_3$ even, and $|l_1-l_2| \leq l_3 \leq l_1+l_2$ \cite{khersonskii1988quantum}. The $3-j$ symbols have many known symmetries and recurrence relations but we will not state them here as they are not relevant for this section, however, reference \onlinecite{khersonskii1988quantum} contains the details for those interested. 

The usual approach to calculating the Gaunt coefficients relies on using the recurrence relations of the Wigner $3-j$ symbols to compute them recursively, which is what has been done in this work. The algorithm that is used is called the Schulten-Gordon-Cruzan algorithm \cite{Sebilleau_1998} and it relies on the recurrence relation 
\begin{align}
\begin{pmatrix} l_1 & l_2 & l_3-1 \\ m_1 & m_2 & m_3 \end{pmatrix} &= -\frac{l_3A(l_3+1)}{(l_3+1)A(l_3)}\begin{pmatrix} l_1 & l_2 & l_3+1 \\ m_1 & m_2 & m_3 \end{pmatrix} \nonumber\\[1.5ex]
&\qquad\qquad\qquad- \frac{B(l_3)}{(l_3+1)A(l_3)}\begin{pmatrix} l_1 & l_2 & l_3 \\ m_1 & m_2 & m_3 \end{pmatrix}\label{sgc_recur}
\end{align}
where
\begin{align}
A(l_3) &= \sqrt{l_3^2-(l_1-l_2)^2}\sqrt{(l_1+l_2+1)^2-l_3^2}\sqrt{l_3^2-m_3^2}\nonumber \\[1.5ex]
B(l_3) &= -(2l_3+1)[l_1(l_1+1)m_3-l_2(l_2+1)m_3-l_3(l_3+1)(m_2-m_1)]\,.
\end{align}
When $m_1=m_2=m_3=0$, the $3-j$ symbols are only non-zero for even $l_1+l_2+l_3$ (third selection rule) and $B(l_3) = 0$, so the second term on the right hand side of eqn. \ref{sgc_recur} vanishes and we shift $l_3$ down by 1 to arrive at the expression
\begin{align}
\begin{pmatrix} l_1 & l_2 & l_3-2 \\ 0 & 0 & 0 \end{pmatrix} = \frac{K(l_3)}{K(l_3-1)}\begin{pmatrix} l_1 & l_2 & l_3 \\ 0 & 0 & 0 \end{pmatrix} \label{sgc_recur_0}
\end{align}
where
\begin{align}
K(l_3) &= \sqrt{l_3^2-(l_1-l_2)^2}\sqrt{(l_1+l_2+1)^2-l_3^2}\,.
\end{align}
The algorithm then works as follows. A particular $l_3$ value is chosen and the recurrences eqn. \ref{sgc_recur} and eqn. \ref{sgc_recur_0} work their way downward starting at $l_3=l_1+l_2$, which has a known form given by 
\begin{align}
\begin{pmatrix} l_1 & l_2 & l_1+l_2 \\ m_1 & m_2 & m_3 \end{pmatrix} &= (-1)^{l_1-l_2-m_3} \times\nonumber\\[1.5ex]
&\quad\times\sqrt{\frac{(2l_1)!(2l_2)!(l_1+l_2+m_3)!(l_1+l_2-m_3)!}{(2l_1+2l_2+1)!(l_1+m_1)!(l_1-m_1)!(l_2+m_2)!(l_2-m_2)!}}\,,
\end{align}
until they reach the given $l_3$.

Now that the computation of the Gaunt coefficients has been illustrated, it must be adapted for the real spherical harmonics. A brute force approach would be to calculate all $3^3=27$ separate cases. A more elegant method, using ideas from \cite{Homeier_Steinborn1996}, expresses the real spherical harmonics as the unitary transformation of the standard spherical harmonics
\begin{align}
Z^m_{l} (\theta,\phi) = \sum_{m'}U^{m}_{m'}Y^{m'}_{l} (\theta,\phi)
\end{align}
then using the property that $Z^{m\,*}_{l} (\theta,\phi) = Z^m_{l} (\theta,\phi)$, the following condition on the unitary matrix elements must hold $U^{m\,*}_{m'} = (-1)^{m'}U^{m}_{-m'}$, from which reference \cite{Homeier_Steinborn1996} derives the unitary matrix elements to be
\begin{align}
U^{m}_{m'} &= \delta_{|m||m'|}\bigg[\delta_{m0}\delta_{m'0} + \frac{1}{\sqrt{2}}\bigg(\Theta(m)\delta_{mm'}+i\Theta(-m)(-1)^{m'-m}\delta_{mm'} \nonumber\\
&\qquad \qquad \ \ -i\Theta(-m)(-1)^{-m}\delta_{-mm'}+\Theta(m)(-1)^{m'}\delta_{-mm'}\bigg)\bigg]
\end{align}
where $\Theta(x)$ is the Heaviside step function. The Gaunt coefficients for the real spherical harmonics are then expressed as
\begin{align}
\alpha^{mm'm''}_{ll'l''} = \sum_{m_1m_2m_3} \text{Re}\left(U^m_{m_1}U^{m'}_{m_2}U^{m''}_{m_3}\right)G^{m_1m_2m_3}_{ll'l''}\,. \label{realgaunt}
\end{align}
The full algorithm to compute the Gaunt coefficients for the real spherical harmonics then consists of first finding the standard Gaunt coefficients using the Schulten-Gordon-Cruzan algorithm, then computing the unitary matrix $U$, and finally computing the matrix product eqn. \ref{realgaunt}.

\section{KS-DFT Equivalence to the SCFT Model}
\label{AppendixC}

The effective Hamiltonian $H^{\text{eff}}_\mu$ appearing in the single-pair diffusion equation eqn. \ref{diffuse} is the same Hamiltonian that appears in the Kohn-Sham equation, therefore the orbitals $\phi_{\mu i}(\bm{r})$ from this Hamiltonian can be used as a basis set and to solve for the single-pair propagator. The orbitals are defined as
\begin{align}
H^{\text{eff}}_\mu\phi_{\mu i}(\bm{r}) = \left[\varepsilon_\mu\right]_i\phi_{\mu i}(\bm{r}) \ \ \text{where} \ \ \int \mathrm{d}\bm{r}\, \phi_{\mu i}(\bm{r})\phi_{\mu j}(\bm{r}) = \delta_{ij} \label{orb}
\end{align}
so the single-particle propagator is
\begin{align}
q_\mu(\bm{r}, \bm{r}', s) = \sum_{ij} \left[q_\mu(s)\right]_{ij} \phi_{\mu i}(\bm{r})\phi_{\mu j}(\bm{r}')\,.
\end{align}
The diffusion equation is then
\begin{align}
\frac{\partial q_\mu(\bm{r}, \bm{r}', s)}{\partial s} &= \sum_{ij}\frac{\partial \left[q_\mu(s)\right]_{ij}}{\partial s}\phi_{\mu i}(\bm{r})\phi_{\mu j}(\bm{r}') = -H^{\text{eff}}_\mu q_\mu(\bm{r}, \bm{r}', s) \nonumber \\[1.5ex]
&= -\sum_{ij}\left[q_\mu(s)\right]_{ij} \left(H^{\text{eff}}_\mu \phi_{\mu i}(\bm{r})\right)\phi_{\mu j}(\bm{r}') = -\sum_{ij}\left[q_\mu(s)\right]_{ij} \left[\varepsilon_\mu\right]_i \phi_{\mu i}(\bm{r})\phi_{\mu j}(\bm{r}')\,.
\end{align}  
Multiplying both sides with $\phi_{\mu k}(\bm{r})\phi_{\mu l}(\bm{r}')$, integrating over $\bm{r}$ and $\bm{r}'$, and using the orthogonality relation eqn. \ref{orb} then gives
\begin{align}
\frac{\partial \left[q_\mu(s)\right]_{kl}}{\partial s} = -\left[q_\mu(s)\right]_{kl} \left[\varepsilon_\mu\right]_k \ \ \rightarrow \ \ \left[q_\mu(s)\right]_{kl} = \delta_{kl} e^{-\left[\varepsilon_\mu\right]_k s}
\end{align}
so
\begin{align}
q_\mu(\bm{r}, \bm{r}', s) = \sum_{i} e^{-\left[\varepsilon_\mu\right]_i s}\phi_{\mu i}(\bm{r})\phi_{\mu i}(\bm{r}')\,.
\end{align}
The pair densities are then given by
\begin{align}
n_\mu(\bm{r}, \beta) =  \frac{N_\mu}{Q_\mu[w](\beta)}q_\mu(\bm{r}, \bm{r}, \beta) = \frac{N_\mu}{Q_\mu[w](\beta)}\sum_{i} e^{-\left[\varepsilon_\mu\right]_i s}|\phi_{\mu i}(\bm{r})|^2\,, \label{pair_orb}
\end{align}
which is exactly the expression for the density in finite-temperature KS-DFT. Clearly eqn. \ref{pair_orb} does not have one-to-one correspondence with the squared modulus of Kohn-Sham orbitals, therefore we should not necessarily expect the pair densities to reproduce the profiles of individual squared moduli of orbitals.


\bibliographystyle{plain}
\cleardoublepage 
\phantomsection  

\bibliography{Angular_Paper}

\nocite{*}

\end{document}